# The discernible and hidden effects of clonality on the genotypic and genetic states of populations: improving our estimation of clonal rates

**Running title:** Estimating clonal rates: an unreachable star?

Solenn Stoeckel[1*], Barbara Porro[2,3], Sophie Arnaud-Haond[3]

[1]INRA, UMR1349 Institute for Genetics, Environment and Plant Protection, Le Rheu, France.

[2]Institute for Research on Cancer and Aging (IRCAN), CNRS UMR 7284, INSERM U1081, Université de Nice-Sophia-Antipolis, Nice 06107, France.

[3]Laboratory Environnements-Ressources, Ifremer, Bd Jean Monnet BP 171, Sète 34203, France.

**Corresponding author:**

[*]Solenn.Stoeckel@Inra.fr

# Abstract

Partial clonality is widespread across the tree of life, but most population genetics models are designed for exclusively clonal or sexual organisms. This gap hampers our understanding of the influence of clonality on evolutionary trajectories and the interpretation of population genetics data. We performed forward simulations of diploid populations at increasing rates of clonality ($c$), analysed their relationships with genotypic (clonal richness, R, and distribution of clonal sizes, Pareto β) and genetic ($F_{IS}$ and linkage disequilibrium) indices, and tested predictions of $c$ from population genetics data through supervised machine learning. Two complementary behaviours emerged from the probability distributions of genotypic and genetic indices with increasing $c$. While the impact of $c$ on R and Pareto β was easily described by simple mathematical equations, its effects on genetic indices were noticeable only at the highest levels ($c>0.95$). Consequently, genotypic indices allowed reliable estimates of $c$, while genetic descriptors led to poorer performances when $c<0.95$. These results provide clear baseline expectations for genotypic and genetic diversity and dynamics under partial clonality. Worryingly, however, the use of realistic sample sizes to acquire empirical data systematically led to gross underestimates (often of one to two orders of magnitude) of $c$, suggesting that many interpretations hitherto proposed in the literature, mostly based on genotypic richness, should be reappraised. We propose future avenues to derive realistic confidence intervals for $c$ and show that, although still approximate, a supervised learning method would greatly improve the estimation of $c$ from population genetics data.

# Introduction

Clonality occurs across the entire tree of life, including all kingdoms of Eukaryota (Avise & Nicholson 2008; Schön *et al.* 2009; Tibayrenc *et al.* 2015). Most, if not all, clonal eukaryotic species alternate between clonal and sexual reproduction at the population scale over a few generations (see Box 1 for the definitions used in this work). This mode of reproduction, called partial clonality (PC), has been reported in a broad range of species, especially primary producers, ecosystem engineers, pathogens and invasive species. Their evolutionary trajectories may thus have major consequences for ecosystem functioning and for human health and development (Schön *et al.* 2009; Yu *et al.* 2016). Challenging environments and the edges of species ranges may also favour populations able to reproduce using PC and putatively populations with higher rates of clonality, emphasising the importance of understanding evolutionary trajectories of PC species when dealing with global changes (Barrett 2016; Barrett 2015; Tibayrenc & Ayala 2012; Yu *et al.* 2016).

---

**Box 1: relevant definitions and concepts**

**Clonal reproduction**
A precise definition of clonal reproduction has been historically used in population genetics, corresponding to "*an individual produces new individuals that are genetically identical to the ancestor at all loci in the genome, except at those sites that have experienced somatic mutations*" (as defined in De Meeus *et al.* 2007, see also Marshall & Weir 1979). This definition implicitly includes reproduction through agametic tissues and apomictic parthenogenesis.

**Partial clonality**
The term partial clonality (PC) refers to the reproductive system of species undergoing both clonal and sexual reproduction through selfing or outcrossing, or both.

**Comment on the use of the terms clonality and asexuality:**
In this work, we favoured the use of *clonality* rather than asexuality due to its etymology. Clone comes from the ancient Greek κλῶνος, referring to a regrowth, a root shoot or, lately, a growing organic extension with new vigour (Liddell *et al.* 1940). Paradoxically, clonality, the initial definition of which mentioned plant regrowth, seems to have been used more by biologists working on animals, while asexuality is more common in the plant literature.

---

> The term asexuality is currently used beyond its initial definitions, applying to all uniparental reproduction with incomplete meiosis schemes, including those occurring beyond prophase I and resulting in higher levels of recombination (Nougue *et al.* 2015). In addition, a societal definition emerged last year: "*a [human] sexual orientation […] not valuing sex or sexual attraction to others enough to pursue it*" (Decker 2015 in Bibr 2018).
>
> **Rate of clonality**
>
> In line with the definition summarised above, the rate of clonality (here, *c*) corresponds to "*…the probability of clonality* versus *sexual reproduction through selfing or outcrossing*" (Marshall & Weir 1979), which in this article corresponds to the ratio of the effective number of descendants produced by clonality to the total effective number of descendants produced in a population (see also Balloux *et al.* 2003; Berg & Lascoux 2000).
>
> When inferred by genotypic and genetic indices in a population sample, this rate is a *proxy* for the idealised number of descendants produced by clonality relative to the total idealised number of descendants produced in the population that would result in the same genotypic and genetic effects.

Despite the prevalence of PC and the potential extent of its consequences at the ecosystem level, the consequences of PC for the evolution of species and the ecological dynamics of their natural populations have been subject to little in-depth theoretical or empirical development (Yonezawa *et al.* 2004). This lack of development makes a substantial number of studies on partially clonal species confusing when analysing population genetics data and interpreting them in terms of demographic and evolutionary dynamics (Avise 2015; Fehrer 2010; Yu *et al.* 2016). Nevertheless, the effects of PC are likely to be extremely important at all spatial and temporal scales. For example, evolutionarily speaking, the ability of a given genotype to persist across generations adds a new target for natural selection, namely, the genotype (Ayala 1998).

Three main knowledge gaps are related to PC: diagnosing it in species where its occurrence is not obviously inferred by classical naturalistic observations (e.g., human pathogens, in contrast to rhizomatic clonal plants); quantifying its extent once a given species is determined to be partially clonal; and understanding its influence on the ecological and evolutionary trajectories of partially clonal species by investigating their population genetics. These gaps have been only partly filled during the past 30 years. The use of molecular markers in a population genetics framework paved the way for easier detection of PC (De Meeûs *et al.* 2006; Halkett *et al.* 2005; Tibayrenc *et al.* 1990) through the discrimination of clonal lineages and detailed analysis of the genotypic and genetic compositions of species suspected of having PC (Bailleul *et al.* 2016, Arnaud-Haond *et al.* 2005; Tibayrenc *et al.* 1990). However, conditions allowing (or not allowing) the detection of PC and its consequences for the trajectories of natural populations over different time scales are likely important yet still poorly understood (Avise 2015; Dia *et al.* 2014; Fehrer 2010; Yu *et al.* 2016). We still face difficulties in inferring the rate of clonal (denoted $c$) versus sexual ($1-c$) reproduction or an approximate but consistent *proxy* for it (*i.e.*, the "level of clonality"). These difficulties prevent access to the empirical information necessary to compare the ecological dynamics and evolutionary trajectories of partially clonal populations living in different environments (McMahon *et al.* 2017). To understand the effect of PC on the fate of natural populations and species, the value of $c$ should first be estimated.

The rate of clonality in natural populations may be estimated by tracking clonal spreads or determining groups of clones. In plants, groups of clones have sometimes been identified at local scales through extremely time-consuming and tedious mark-recapture studies of rhizomes (Eckert 2002; Marbà & Duarte 1998). However, using this method on large spatial scales and for most species exhibiting PC through fragmentation or multiplication at microscopic stages is unrealistic. Therefore, tracking clonal spread or determining groups of clones through population genetics is the only solution for the vast majority of species. Unfortunately, although population genetics studies can illuminate the occurrence of PC in nature, no method has been developed thus far to reliably infer (or at least estimate) such potentially crucial parameters in natural populations using indices gathered through a classical one-time step sampling strategy. Two recently developed methods allow the quantification of rates of clonality in populations genotyped at two time steps. However, they require sampling the population twice at an interval of at least one generation and, more importantly, a comprehensive knowledge of major life history traits, such as generation time, which are seldom available except for well-known macroscopic species for which extensive field data have been collected (Ali *et al.* 2016; Becheler *et al.* 2017).

Most empirical studies thus infer the importance of clonal reproduction in populations using a one-time step sampling strategy to compute the ratio of genotypes to the number of sampling units (genotypic richness, $Pd = G/N$) as an estimate of clonal richness. Genotypic richness is often implicitly assumed to have a linear relationship with the rate of sexual reproduction $1 - c$ and to be comparable among natural populations submitted to the same sampling strategy. Theoretical studies have shown the strong influence of high clonality rates (c>0.95) alone on parameters such as $F_{IS}$ and linkage disequilibrium (*LD*) (Balloux *et al.* 2003; De Meeûs *et al.* 2006; Navascués *et al.* 2010) but no noticeable departure from expectations under purely sexual reproduction at lower rates of clonality. However, more recent mathematical developments have shown that the distribution of $F_{IS}$ is wider at high clonality rates but is actually affected at all clonality rates (Stoeckel & Masson, 2014), depending on the strength of departure from equilibrium (Reichel *et al.* 2016).

This research led to a present-day paradox in the literature on PC. Many populations exhibit average or elevated genotypic diversity, leading several authors to conclude that these populations exhibit a high incidence of sexual reproduction, whereas in the same studies, consistent departure from Hardy-Weinberg equilibrium (HWE), when reported (which is much rarer), would instead have led them to conclude that the populations exhibit a negligible occurrence of sexual recombination versus clonal reproduction (e.g., Orantes *et al.* 2012; Villate *et al.* 2010). This paradox is seldom obvious because $F_{IS}$ values are often not reported or, if reported, are not interpreted in relation to clonality. In any case, part of this paradox may lie in the pervasive effect of sampling on the estimation of genotypic richness (Arnaud-Haond *et al.* 2007; Gorospe *et al.* 2015). These two studies demonstrated this worrying effect by using two empirical datasets (of seagrasses and corals) where the true rates of clonality were unknown; assessing the order of magnitude of these rates thus requires further investigation.

*A roadmap to fill the gaps*

Given the current state of knowledge, the characterisation of the genotypic (based on groups of clones) and genetic (based on allele and genotype frequencies at loci) compositions of populations is both the target and the *proxy* of population genetics studies aiming to understand the influence of clonal reproduction on the dynamics and evolution of natural populations. Reconciling the effects of PC on both the genotypic and genetic compositions of populations in a robust theoretical framework is thus necessary to illuminate the concomitant changes in their respective estimators depending on the rate of clonality.

Here, we propose a simulation-based exploratory approach both to enhance our understanding of the consequences of clonality and to improve our ability to reliably assess its rate within natural populations. We aim to provide the first exploration of the effect of increasing $c$ on the genotypic and genetic compositions of populations to provide baseline expectations for the composition of natural populations depending on the extent of clonality. We used comprehensive forward individual-based simulations to obtain the theoretical distribution of genotypic (genotypic richness and size distribution of lineages) and genetic (departure from HWE and LD) parameters describing the population composition at increasing rates of clonality from 0 to 1, including all populations with some clonal reproduction, hereafter denoted PC for brevity, ranging from partial ($0<c<1$) to strict clonality ($c=1$), and populations with solely sexual reproduction ($c=0$), hereafter denoted sexual populations. We explored the temporal evolution of these populations along trajectories towards equilibrium and under various levels of genetic drift (population sizes spanning three orders of magnitude). To move from insights about the expected effects of PC on natural populations towards more reproducible and formalised arguments, we assessed the signature of PC in the genotypic and genetic index distributions using a classical and robust Bayesian supervised learning method. This method allowed the selection of descriptors that were more clearly affected to in turn develop sound estimates of the extent of clonality. Finally, we tested the robustness of the method in examining the influence of sample size on the accuracy of estimates and proposed further improvements based on realistic sample sizes.

## Materials and methods

*Approach*

PC is empirically known to affect genotypic and genetic descriptors commonly used in population genetics studies (Halkett *et al.* 2005): 1) the number of different genotypes per population, as characterised by the genotypic richness indices *R* and Pareto *β* (Arnaud-Haond *et al.* 2007), and two genetic indices, namely, 2) the inbreeding coefficient $F_{IS}$ and its moments (Balloux *et al.* 2003; Stoeckel & Masson 2014) and 3) the LD index (Navascués *et al.* 2010). To date, no analytical formalisation has been developed to predict the theoretical probability distributions of these descriptors under varying rates of clonality. We thus used simulations to i) synthesise the effects of varying rates of clonality on the ranges and dynamics of these genotypic and genetic descriptors, ii) assess whether these descriptors actually provide the ability to discriminate and quantify rates of clonality using a classic supervised learning method, and iii) determine which descriptors best account for specific ranges of rates of clonality, with the aim of providing recommendations for future analyses and interpretations.

*Simulations*

Theoretical results were obtained using forward individual-based simulations run over $10^4$ non-overlapping generations to reach quasi-stationary distributions of both genotypic and genetic diversity. In the initial generations, alleles at all neutral loci were randomly drawn from a uniform distribution (*i.e.*, maximum genetic diversity merged at random within individuals). In these simulations, all diploid individuals lived in constant finite-sized populations.

Each population produced the next generation using clonal or panmictic sexual reproduction following a fixed rate of clonality. All hermaphrodite individuals in each generation had identical probabilities of being parents, in both clonal and sexual events. The probability of an individual parent being drawn i) to birth a clonal descendant and ii) to sire half a sexual descendant followed a Bernoulli scheme, with respective probabilities $P(clonal\ parent) = \frac{c}{N}$ and $P(sexual\ parent) = \frac{1-c}{2N}$, where $N$ is the population size. In clonal reproduction, new independent individuals were produced as full genetic copies of their only parent, with somatic mutations occurring at a fixed rate of $10^{-6}$ mutations per generation per locus. This choice was driven by the high end of estimates of DNA polymerase mutations ranging between $10^{-8}$ and $10^{-9}$ bp/generation (McCulloch & Kunkel 2008), which for a locus of 100 to 1000 base pairs would imply a mutation rate of $10^{-5}$ to $10^{-7}$. In panmictic sexual reproduction, new independent individuals descended from two parents chosen at random within the previous generation, from which the individuals inherited half their genomes and mutated at a rate of $10^{-3}$ mutations per generation per locus, following estimated mutation rates for sexual eukaryotes ranging from $10^{-4}$ to $10^{-7}$ and $10^{-2}$ to $10^{-5}$ across generations for single-nucleotide polymorphisms (SNPs) and microsatellites, respectively (Payseur & Cutter 2006). Genomes were coded as 100 independent loci. Alleles mutated following a K-allele mutation (KAM) model (Putman & Carbone 2014; Weir & Cockerham 1984), which has the advantage of simulating the behaviour of both microsatellites and SNPs well and which best approximates the "disturbing factor of gene frequencies" (in the sense of Wright 1931) in finite-sized populations. Mutating alleles in both clonal and sexual reproduction were drawn at random from the respective pools of clonal and sexual offspring. In simulations, the clonality rate, genetic drift and mutation rate were applied homogeneously across generations and loci.

To understand the effect of clonality on population genetics indices, we ran simulations with varying population sizes ($N= 10^3$, $10^4$ and $10^5$, to be studied with arbitrarily fixed mutation rates), rates of clonality ($c$=0, 0.1, 0.2, 0.3, 0.4, 0.5, 0.6, 0.7, 0.8, 0.9, 0.99 and 1), and numbers of generations elapsed since the initial population (generations=10, 100, 500, 1000, 5000, and 10000). At each time step, indices were examined for the whole population (all $N$ genotypes considered) as well as for subsamples without replacement of different sizes ($n$=10, 20, 30, 50, 100, 200, 500, and 1000 and when population sizes allowed, $n$=5000, 10000, 50000, and 100000).

Each scenario was run 100 times and characterised by a set of parameters ($N$, $c$). When subsampling populations, we performed 10 independent resamplings of each generation and sample size, resulting in 1000 independent data points per set of parameters for each sample size.

*Genotypic and genetic descriptors*

To account for the genotypic composition and genetic state of populations, we computed two indices describing the number and distribution of genotypes (genotypic richness $R$ and slope of the size distribution of lineages Pareto $\beta$, Arnaud-Haond *et al.* 2007) and two genetic descriptors referring to intra-individual genetic variation (as the first four moments of the inbreeding coefficient $F_{IS}$ distribution; Stoeckel & Masson 2014) and LD (as the summarised unbiased multi-locus LD $\bar{r_D}$; Agapow & Burt 2001).

*Genotypic richness*

The $R$ index of clonal diversity (Dorken & Eckert 2001) was defined as follows:

$R = (G - 1)/(N - 1)$

where $G$ is the number of distinct genotypes (genets) and $N$ is the number of genotyped individuals.

*Size distribution of lineages*

The parameter Pareto $\beta$ describes the slope of the power-law inverse cumulative distribution of the size of lineages (Arnaud-Haond *et al.* 2007):

$$N_{\geq X} = a.X^{-\beta}$$

where $N_{\geq X}$ is the number of sampled ramets belonging to genets containing $X$ or more ramets in the sample of the population studied, and the parameters $a$ and $\beta$ are fitted by regression analysis.

*Genetic variance apportionment*

The Wright (1921, 1969) inbreeding coefficient $F_{IS}$ accounts for intra-individual genetic variation as a departure from Hardy-Weinberg assumptions of the genotyped populations. We computed one $F_{IS}$ value per population and per locus as $F_{IS_l} = \frac{Q_{w,l} - Q_{b,l}}{1 - Q_{b,l}}$, where $Q_{w,l}$ is the population probability that two homologous alleles taken within individuals are identical at locus $l$, and $Q_{b,l}$ is the population probability that two homologous alleles taken between different individuals are identical at locus $l$. We computed the first four moments of the empirical $F_{IS}$ distribution obtained from the 10000 independent $F_{IS}$ values per scenario (100 independent loci x 100 replicated simulations), respectively noted $Mean[F_{IS}], Var[F_{IS}], Skew[F_{IS}]$ and $Kurt[F_{IS}]$.

*Linkage disequilibrium between loci*

LD was studied using $\bar{r}_d$ (Agapow & Burt 2001). The mean correlation coefficient (*r*) of genetic distance (*d*) between unordered alleles at $n$ loci ranged from 0 to 1. This metric has the advantage of limiting the dependency of the correlation coefficient on the number of alleles and loci and is well suited to studies of partially clonal populations.

$$\bar{r}_d = \frac{V_D - \sum_{j=1}^{j=n} var_j}{2 \sum_{j=1}^{j=n} \sum_{k>j}^{k=n} \sqrt{var_j . var_k}}$$

with $V_D = \frac{\sum_{a,b \neq a}^{v} D_{a,b}^2 - \frac{(\sum D_{a,b})^2}{v}}{v}$ and $var_j = \frac{\sum d^2 - \frac{(\sum d)^2}{v}}{v}$

where $D$ is the number of loci at which two individuals, $a$ and $b$, differ (the genetic distance between two individuals over all their loci), $d$ is the number of different alleles between two individuals at locus $j$ (for diploids, $d$ can be 0, 1 or 2), and $v$ is the number of unique possible pairs of individuals $a$ and $b$, where $b \neq a$, within a population.

*Genotypic descriptors as empirical functions of the rate of clonality*

To assess the relation between $c$ and the genotypic descriptors, we explored the mean results of simulations as a function of $c$. Depending on the shape of the curves obtained with simulated data, we tested the fit with basic functions (for example, simple sigmoids and exponentially decreasing distributions) as well as with sigmoid and parabolic curves. To assess the accuracy of our empirically inferred formula to describe the relationships, we computed the mean absolute error (MAE) and the root-mean-square deviation (RMSD) between pseudo-observed simulated values and fitted formulae. These two deviation measures aggregate the magnitudes of the errors of predictions into a single measure of predictive accuracy. This measure represents the mean deviation of predicted values with respect to the observed values and has the advantage of sharing the same units as the model variable under evaluation. Lower deviation measures indicate higher accuracy of an analytical formula in the prediction of data. These measures must be interpreted at the same scale as the mean value of the studied parameter (Piñeiro *et al.* 2008).

$$MAE = \frac{1}{n} \cdot \sum_{1}^{n} |y_s - y_f|$$

$$RMSD = \sqrt{\frac{1}{n} \cdot \sum_{1}^{n} (y_s - y_f)^2}$$

where *n* is the number of pseudo-observed simulations per scenario, $y_s$ is the simulated value of the genotypic descriptor under consideration, and $y_f$ is the calculated value of the genotypic descriptor using the fitted formula.

*Identifiable signals in genotypic and genetic descriptors, inferences and machine learning*

Our second objective was to test for the ability of genotypic and genetic descriptors to estimate specific rates of clonality. These descriptors were commonly used in previous studies to roughly assess the importance of clonality in determining population reproductive modes, but no theoretical development has demonstrated the existence of identifiable signals allowing such descriptors to be used as key parameters with which to estimate rates of clonality. To assess the existence of identifiable signals in these descriptors and demonstrate their potential usefulness in inferring rates of clonality for one episode of genotyping, we used the results obtained from the simulations as classifiers to train a Bayesian supervised learning algorithm. We used the simulation results to compute the approximate nonparametric probability distributions of the genotypic and genetic descriptors (*i.e.*, the seven *features* $\varphi_7 = [R, \beta_p, \overline{r_d}, Mean[F_{IS}], Var[F_{IS}], Skew[F_{IS}], Kurt[F_{IS}]]$) with combinations of Gaussian kernels under known rates of clonality, resulting in a *classifier* with 12 classes (one class for each rate of clonality to be inferred: c=0, 0.1, 0.2, 0.3, 0.4, 0.5, 0.6, 0.7, 0.8, 0.9, 0.99 and 1), hereafter referred to as $C_{12}$.

$$L(\varphi_7|C_{12}) = L(R, \beta_p, \overline{r_d}, Mean[F_{IS}], Var[F_{IS}], Skew[F_{IS}], Kurt[F_{IS}]|N, c, \mu)$$

Provided that dependencies between the seven genotypic and genetic descriptors are evenly distributed or cancel each other out or that their distributions sufficiently segregate over their means per class, we can approximate the joint probability model using the conditional independence between features (Hand & Yu 2001; Webb *et al.* 2005; Zhang 2004). The posterior probability of the *i*[th] class, given that the seven measured features are known, can be expressed as the product of the seven likelihoods of each feature weighted by the prior probability of the class.

$$P(C_i|\varphi_7) = p(C_i) \cdot \prod_{j=1}^{7} L(\varphi_j|C_i)$$

From this joint posterior probability, we identified the maximum *a posteriori* (*MAP*) to discern the class ("rate of clonality" and "population size" pair) most likely to explain the measured features.

$$MAP = \underset{i \in \{1,\dots,12\}}{\mathrm{argmax}} \left[ p(C_i) \cdot \prod_{j=1}^{7} L(\varphi_j|C_i) \right]$$

We assumed a uniform distribution prior, *i.e.*, equiprobability for each class $p(C_i) = 1/12$, to place the algorithm in an initial state of complete ignorance of the likely values that the two parameters might take.

We built training and test databases of 100 and 30 replicates per *rate of clonality and population size* pair, respectively. We explored by cross-validation whether there were enough identifiable signals in the features of our classifier $C_{12}$ to infer the true rates of clonality with known values of only population genotypic ($R, \beta_p$) and genetic ($F_{IS}, \overline{r_d}$) indices alone and in combination. Posterior distributions of the thirty test pseudo-observed datasets per *rate of clonality* and *population size* pair were combined to plot the results.

**Results**

We first explored the results at equilibrium to understand the influence of clonality on $R$, Pareto $\beta$, LD measured as $\overline{r}_d$, and the mean, variance, skewness and kurtosis of $F_{IS}$ at three population sizes ($N=10^5$, Figure 1; $N=10^3$, Figure S1.a and $N=10^4$, Figure S1.b) and then examined the evolutionary dynamics of the parameters over generations to determine the effect of clonality at different time steps and quantify the time needed to converge towards stationary values (Figure 2, Figure S2). We assessed which genotypic and genetic parameters produced the most identifiable signal, allowing accurate inferences (Figure 3, Figures S3 and S4). Finally, we approached the issue of sampling strategy to determine its effects on the accuracy of estimates for datasets obtained from natural populations (Figure 4, Figure S2).

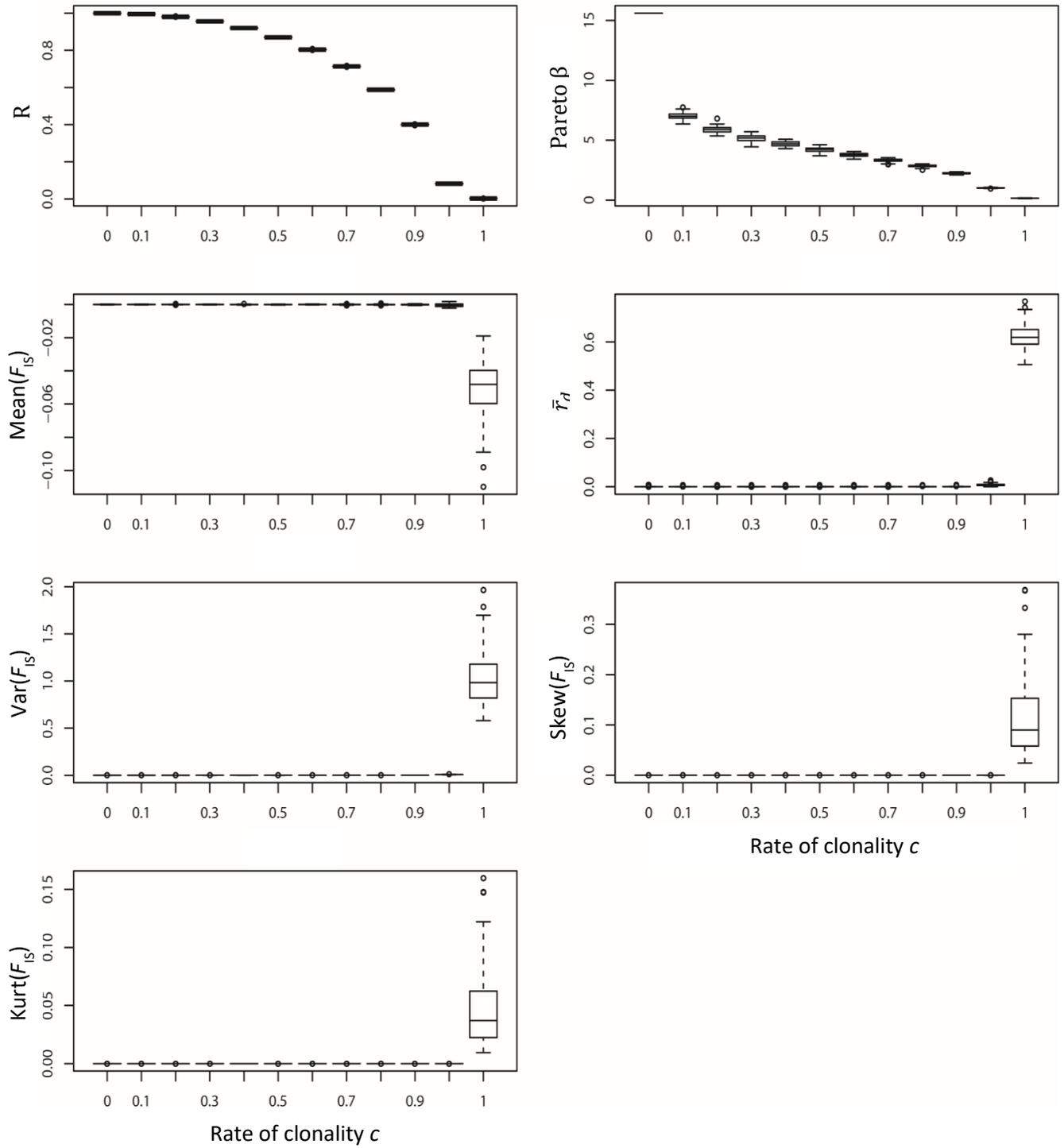

**Figure 1.** Distribution of parameters explored at equilibrium ($10^4$ generations of quantitatively homogeneous evolution since the initial random population) as a function of rates of clonality, *c*, at population size N=$10^5$ (for N=$10^3$ and N=$10^4$, see supplementary Figure S1): genotypic parameters: (a) R and (b) Pareto β; and genetic parameters: (c) $F_{IS}$ mean, (d) $F_{IS}$ variance, (e) $F_{IS}$ skewness, (f) $F_{IS}$ kurtosis and (g) linkage disequilibrium. measured as $\bar{r}_d$. The X-axis is linear from *c*=0 to *c*=0.9 and then non-linear for the last two boxes at *c*=0.99 and *c*=1.

*Genotypic richness and the distribution of clonal size at equilibrium under an increasing rate of clonality*

In terms of genotypic diversity, our results showed a clear, progressive, and even stepwise decrease with increasing rates of clonality (Figure 1, Figure S1).

When genotyping the entire population, the relationship between *R* and *c* (Figure 3) does not follow a linear trend, such as $R = 1 - c$, as might have been assumed in some previous studies. The relationship is best modelled by $R = \sqrt{1 - c^2}$ (*N*=100000: *MAE*=0.011 and *RMSD*= 0.020 for $\bar{R} = 0.69$). The same equation fits the simulation results regardless of the population size, with slightly larger deviations at smaller population sizes, as expected with an increasing strength of genetic drift (*N*=10000: *MAE*=0.013 and *RMSD*= 0.021 for $\bar{R} = 0.70$; *N*=1000: *MAE*=0.029 and *RMSD*= 0.041 for $\bar{R} = 0.71$), but still providing an accurate approximation.

The curve describing the evolution of the parameter Pareto *β* in the power-law distribution of clonal sizes depending on the rate of clonality shows a slightly more complex pattern. The curve has the typical shape of a sum of two sigmoid curves with three sub-domains delimited by two inflection points (Figure 1). Very low levels of clonality (0<*c*<0.1) lead to maximum Pareto *β*-values, which depend on the population size (approximately 8 for *N*=100 individuals to 15 for *N*=100000 individuals). For these distinct initial values, the curves show an extremely similar shape regardless of population size, with a marked sigmoid shape of Pareto *β*-values declining from approximately 8 (value corresponding to high richness and evenness; Arnaud-Haond *et al.* 2007) at *c*=0.1 to nearly 0 for *c*=1. Interestingly, the value *β*=2 is reached for clonal rates of approximately 0.8 to 0.9 for all population sizes. Between clonal rates of 0.2 and 0.9, the decline in $\beta$ is nearly linear and flat for all population sizes. For $N = 100000$, the sum of two fitted sigmoid curves produces the following equation:

$$\beta = \frac{337335}{1+e^{16\times(c+0.65)}} + \frac{5}{1+e^{6.8\times(c-0.80)}} \text{ } (MAE\text{=}0.30 \text{ and } RMSD\text{= }0.40 \text{ for } \bar{\beta} = 4.66);$$

for $N = 10000$,

$$\beta = \frac{506607}{1+e^{9.8\times(c+1.12)}} + \frac{4}{1+e^{8.3\times(c-0.81)}} \text{ } (MAE\text{=}0.27 \text{ and } RMSD\text{= }0.36 \text{ for } \bar{\beta} = 3.94); \text{ and}$$

for $N = 1000$,

$$\beta = \frac{5.6}{1+e^{5\times(c-0.58)}} + \frac{3.8}{1+e^{50\times(c-0.19)}} \text{ } (MAE\text{=}0.32 \text{ and } RMSD\text{= }0.40 \text{ for } \bar{\beta} = 3.63).$$

*Evolution of the genetic composition of populations under an increasing rate of clonality*

In contrast to the genotypic results but in agreement with previous studies on populations at equilibrium with a realistic low mutation rate (Balloux *et al.* 2003; Navascués *et al.* 2010; Stoeckel & Masson 2014), all mean genetic indices are nearly unaffected until the rate of clonality reaches 0.95 (Figure 1, Figure S1). In fact, only a slightly larger variance, exemplified at smaller population sizes, can be observed at $c$=0.9 for $F_{IS}$ and its moments and for $\bar{r}_d$. The effects of $c$ on genetic parameters are thus limited to extreme $c$ values.

When the rate of clonality reaches 0.99, values of $F_{IS}$ are slightly negative, a situation perceptible mostly in a small population ($N$=1000, Figure S1.a). The three deeper moments of $F_{IS}$ distributions, however, show values strongly departing from zero. At this highly elevated rate of clonality, LD very slightly departs from 0 for large population sizes ($N\geq10000$) and then shows extreme values at $c$=1 ($\bar{r}_d$ of approximately 0.8 and 0.6 for $N$=10000 and 100000, respectively). Interestingly, very small populations show a modest (mean value of approximately 0.06) but noticeable departure from 0 as well as large variance at $c$=0.99, whereas $\bar{r}_d$ returns to zero (in fact to a very slightly negative value) at $c$=1, with a more limited variance. This unexpected behaviour occurs because clonality, by increasing the number of generations to reach the genotype frequencies expected under Hardy-Weinberg assumptions, allows genetic drift to control the dynamics of genetic diversity (Reichel *et al.* 2016; Rouger *et al.* 2016). In strictly clonal small populations containing 1000 individuals, the number of genets ranges from a minimum of 91 to a maximum of 102, with a median of 97. These populations are dominated by one main multi-locus genotype (MLG), and the remaining MLGs are scarcely represented (two or three copies each), appearing as derived from the main one only through somatic mutations. These populations thus consist of the same multi-locus lineage (MLL) characterising the genet, *i.e.*, the ensemble of ramets issued from the same event of sexual reproduction (Arnaud-Haond *et al.* 2007). MLGs mostly diverge from each other by 1 to a maximum of 16 alleles (median=2) over a total of 200 alleles per MLG. The $\bar{r}_d$ values are thus driven by the random association of the few alleles recently appearing by mutation in an overdominant clonal lineage fixed by genetic drift. MLGs differing by very few loci imply that $V_D$ tends to zero and that $var_j$ tends to non-zero positive values at each locus $j$.

*Evolution of parameters when moving towards equilibrium*

For $c \leq 0.99$, the maximum number of generations required to reach the stationary mean value of the genotypic descriptor was several tens to several hundreds, whereas for $c>0.99$, the convergence time far exceeded 300 generations (Figure 2, Figure S2). For the genetic indices, a nearly stable mean value was observed beginning in generation 100, with very small fluctuations. Strictly clonal populations ($c$=1) differed: their genotypic parameters also reached a steady value early, but their genetic parameters continued to evolve for 10000 or more generations. These simulation results are in line with results obtained with mathematical assessments (Reichel *et al.* 2016): the farther the population is from its equilibrium, the faster it converges towards the equilibrium values. The trajectory then slows down as the values approach those expected at equilibrium. Mathematical analysis predicted that the equilibrium values would be asymptotically reached after a maximum convergence time that depended on the relative strength of clonality, genetic drift and mutation.

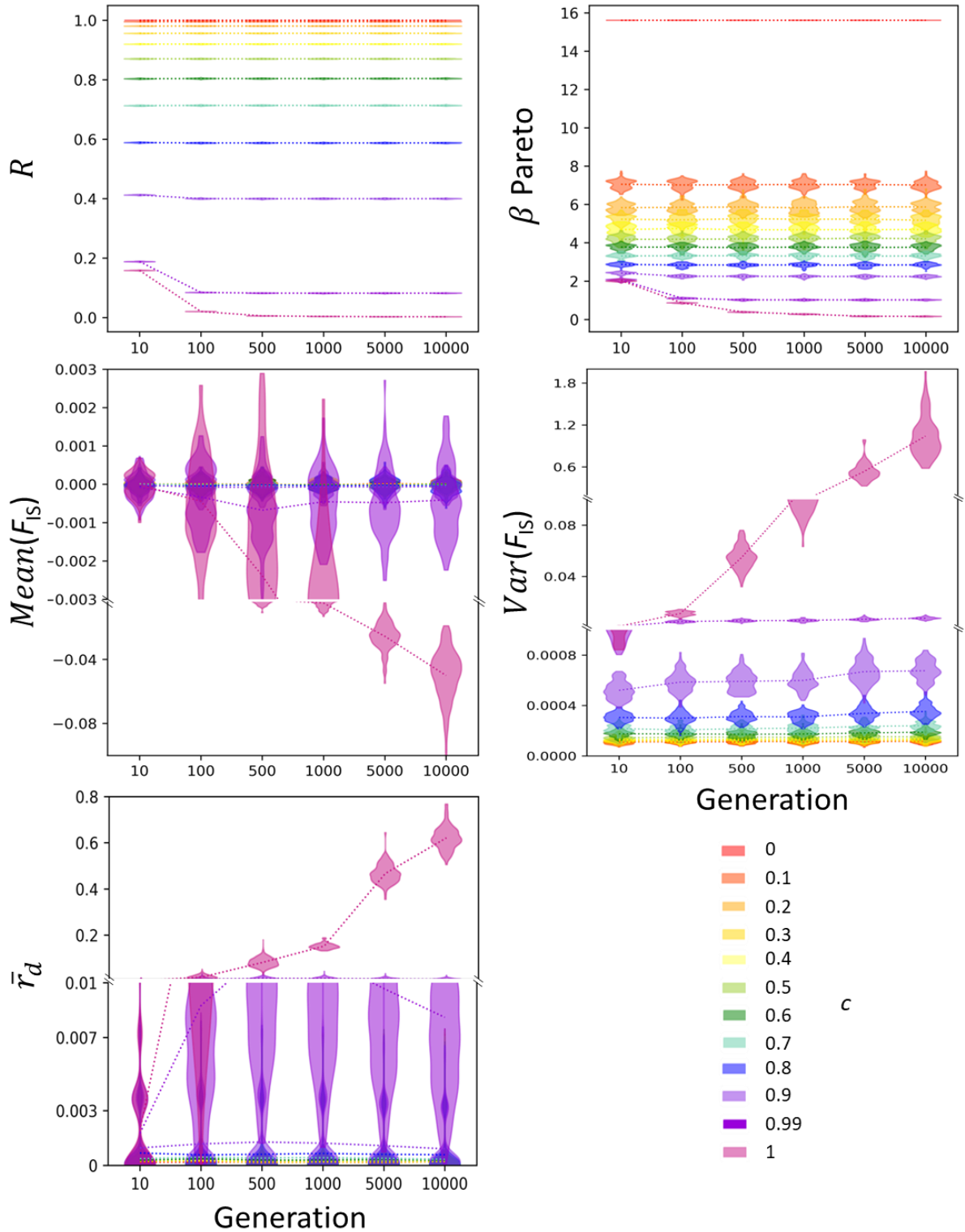

**Figure 2.** Temporal evolution of each parameter at a population size of $10^5$ individuals per generation, as a function of the number of generations elapsed from a fully random population at generation 0. Genotypic parameters: (a) R and (b) Pareto β; genetic parameters: (c) FIS mean, (d) FIS variance, (e) linkage disequilibrium measured as r̄d. For smaller population sizes (N=$10^3$

and N=$10^4$), see supplementary Figure S2 and S2.b. Caution regarding interpretation: all x-axes are non-linear, and the y-axis for $\bar{r}_d$ and the mean and variance of the $F_{IS}$ distributions present one to two changes in scaling.

*Identifiability of genotypic and genetic signals using machine learning*

Supervised machine learning, as expected, delivers estimates consistent with the description of parameter evolution with increasing $c$ when all genotypes in the population are known (Figure 3, Figure S3). Genotypic indices allow a reasonable estimate of $c$ throughout its range, while genetic parameters allow such precision only for very high values. Genotypic parameters evolve gradually with high accuracy of the estimated $c$ based on $R$ and a slightly wider but still rather precise distribution when based on an intermediate Pareto $\beta$. In contrast, but logically (as the mean values of genetic parameters are nearly unaffected by increasing clonality until extreme rates are reached; Figure 1, Figure S1), machine learning produced a wide distribution of estimates around simulated values of $c$ up to $c=0.6$ for $F_{IS}$ and $c=0.9$ for $\bar{r}_d$. This distribution, however, is not entirely flat, and although $c$ estimates are poor at modest rates of clonality, they become precise near values of $c$ between 0.7 and 0.99.

Based on supervised machine learning, the variance in $F_{IS}$ was the most identifiable signal among the studied genetic parameters (Figure S4). The mean and variance of $F_{IS}$ contain more identifiable signals than $\bar{r}_d$ in the range of $0 < c < 0.9$. The mean and variance of $F_{IS}$ values even show rather accurate inferred rates of clonality from $c=0.7$ to $c=1$. The variance in $F_{IS}$ showed the best ability to quantitatively infer $c<0.5$ but produced an error of $\pm 0.3$. Using all moment values of $F_{IS}$ and $\bar{r}_d$, the supervised learning algorithm groups the strength of these parameters, increasing the precision to quantitatively infer $c$. Rates of clonality from $c=0.7$ to $c=1$ were inferred with no error; from $c=0.4$ to $c=0.7$, with low error ($\pm 0.1$); and from $c=0$ to $c=0.4$, with larger errors ($\pm 0.3$).

Taken together, the genotypic and genetic parameters thus complement each other to properly estimate $c$, with the first allowing very precise estimates of $c$ up to 0.95, where the latter become useful and precise. The combination of genotypic and genetic parameters should thus be considered to precisely estimate the whole range of possible rates of clonality in natural populations, although genotypic parameters *a priori* appear to be the most important to retain across the widest range of possible *c* values. Theoretically, a combination of *R* and (variance in) $F_{IS}$ would be best for obtaining a good estimate of *c* for any natural population when no *a priori* information on its extent is available.

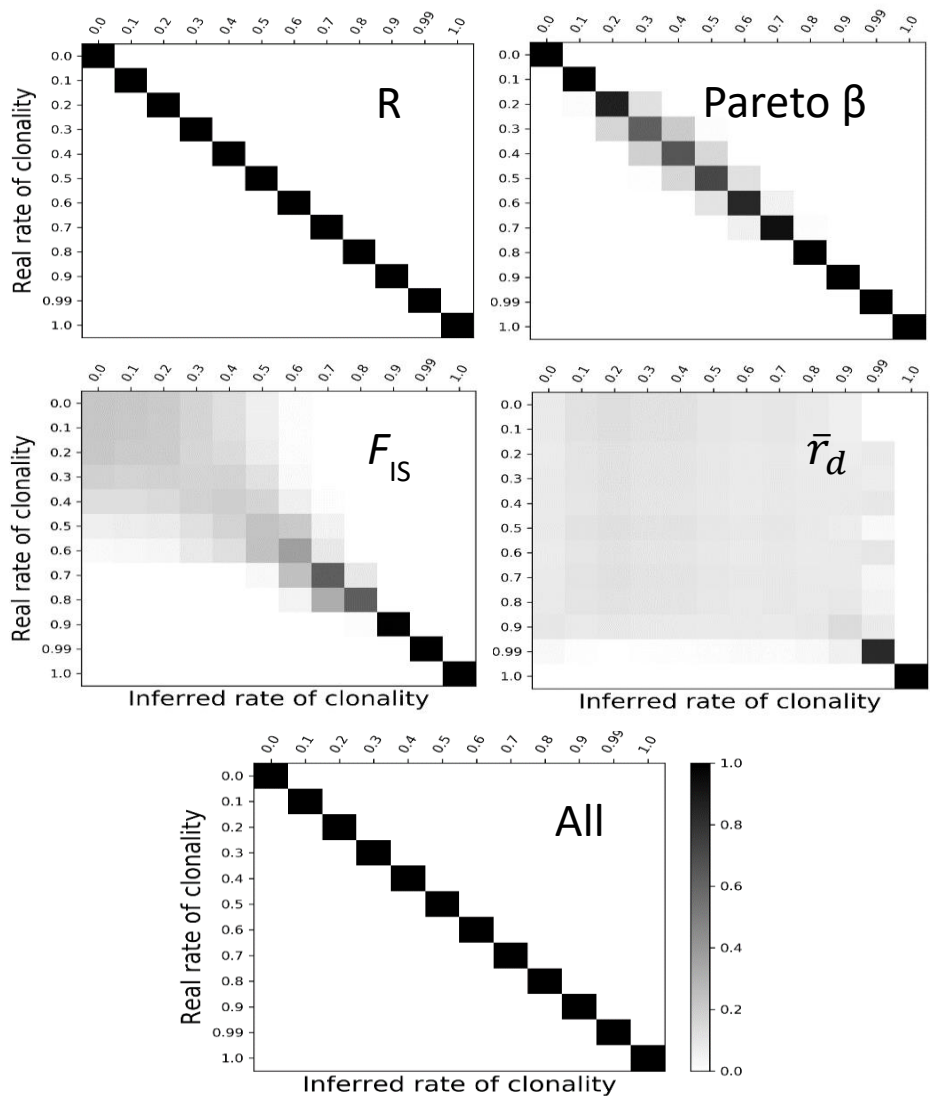

**Figure 3.** Machine learning inferences of *c* at $N=10^5$ and for each parameter used for inference: genotypic parameters (a) R and (b) Pareto β and genetic parameters (c) $F_{IS}$ and (d) $\bar{r}_d$, as well as (e) the combination of all four parameters. The inferred values are plotted against the simulated values, with the density gradient from black to light grey indicating the most to least likely/probable.

*Subsampling*

The inference method described above assumes that all individuals from the *in silico* population have been sampled and genotyped. When subsampling is applied in a realistic manner (*i.e.*, mimicking the subsampling level of most studies in molecular ecology), however, real issues emerge in terms of parameter accuracy and consequent estimates of *c*.

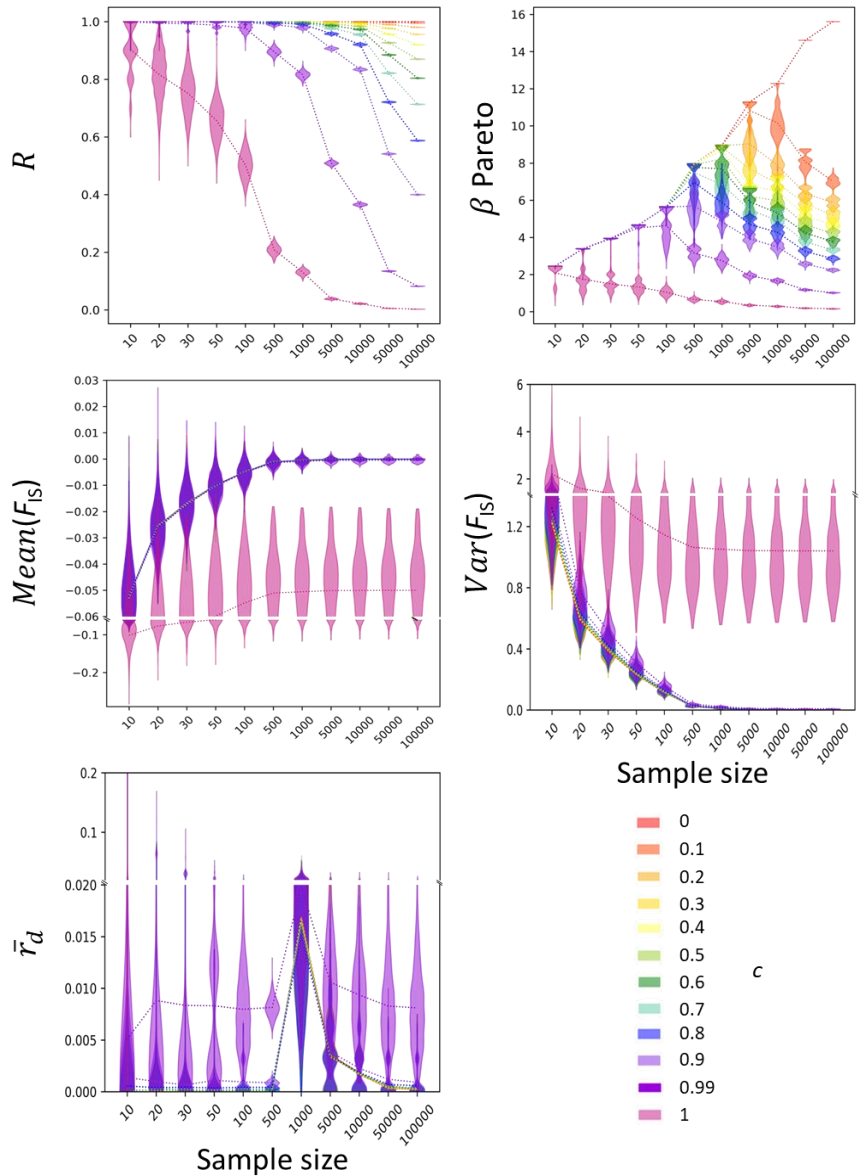

**Figure 4.** Subsampling effects on the distributions of genotypic indices (R and Pareto β) and genetic indices (mean and variance of the $F_{IS}$ distribution and LD measured as $\bar{r}_d$), depending on the sample sizes applied to the dataset, with $N=10^5$ at equilibrium (generation g=10000). For smaller population sizes ($N=10^3$ and $N=10^4$) and for the combined effect of subsampling and a non-equilibrium state, see supplementary Figure S2.a and S2.b. Caution regarding interpretation: all x-axes are non-linear, and the y-axis for $\bar{r}_d$ and the mean and variance of the $F_{IS}$ distributions presents one change in scaling.

Genetic parameters (which proved to be less informative for assessing the clonal rate) were nearly unaffected by realistic sample sizes, whereas genotypic parameters (which were most informative) were considerably overestimated when using realistic sample sizes, leading to a gross underestimate of c from real datasets collected from natural populations. This situation remained nearly unchanged when subsampling was performed before the population reached equilibrium (see Figure S2). The $R$ parameter is so susceptible to sampling bias that a sampling effort of 50 units, consistent with many works studied thus far, including ours, cannot reliably estimate an $R$ value lower than 0.9, with the exception of highly clonal populations ($c$>0.8). A correct and unbiased estimate of R can be achieved only by genotyping the entire population (Figure 4, Figure S2). Interestingly, the variance in $F_{IS}$ computed from samples provided more identifiable signals of rates of clonality at all populations sizes than genotyping all individuals. However, away from equilibrium, the variance in $F_{IS}$ became less informative than that obtained at equilibrium (Figure S2).

## Discussion

This work sheds new light on the precise influence of PC on the genotypic and genetic composition of natural populations, allowing the identification of the most accurate parameters that should theoretically be used to estimate $c$ under a wide range of conditions. Nevertheless, the results also clearly demonstrate that the most useful parameters, namely, the $R$ and Pareto $β$ describing genotypic diversity, are seriously affected by sampling density. This vulnerability raises questions about our ability to detect clonality by sampling large populations only once and serious doubts as to the possibility of quantitatively inferring rates of clonality in large populations based on genotypic diversity alone.

These findings stimulate new interpretations of some published data and perspectives on improvements that are required to further understand the dynamics and evolution of the broad range of species exhibiting PC.

*Parameters most influenced by c and the consequent accuracy of inferences based on these parameters*

The index $R$ (clonal richness) is widely used to assess the level of clonality within natural populations, especially in correlation with environmental drivers, to decipher the impacts of ecological features on PC (McMahon *et al.* 2017). Using entire populations, we empirically formalised the mathematical relationship between $c$ and the genotypic richness indices $R$ and $\beta$ (Figure 1, Figure S1). The relationship with $R$ is not linear (as sometimes seemingly assumed in the literature) but follows $R = \sqrt{1-c^2} \pm \varepsilon$ (with $\varepsilon$ being a small, positive, almost zero random error depending on the strength of genetic drift). Clonal evenness, represented by Pareto $\beta$, is also not a linear function of $c$. Instead, Pareto $\beta$ follows a custom sigmoid curve with three domains (ranging from $c \in [0, 0.15]$, $[0.15, 0.9]$ and $[0.9, 1]$), with the first and last showing a strong decrease in Pareto $\beta$ with increasing $c$. In contrast, in the smooth linear domain ranging from $c\sim0.15$ to $c\sim0.9$, the relationship is almost horizontal, suggesting limited changes in genotypic evenness in populations with balanced amounts of sexual and clonal events.

In contrast, and in agreement with previous findings (Balloux *et al.*, 2003; De Meeûs *et al.*, 2006), the genetic parameters are, on average, largely unaffected below extreme rates of clonality ($c<0.95$). However, the variance in $F_{IS}$ and $\bar{r}_d$ hints at PC and should theoretically allow estimation of its extent under a high prevalence of clonality ($c \geq 0.95$; Figure 1, Figure S1). Clonality acts by releasing the coercive effects of sexuality that constrain and channel the evolutionary trajectories of genotype frequencies towards Hardy-Weinberg proportions, which in turn increases the range of possible values for the genetic indices. This effect results in a broader distribution of genetic indices with a larger variance and unusual shapes, despite nearly unaffected mean values. The effect of clonality on the composition of natural populations is thus expected to be much more pronounced in terms of the genotypic structure, which strongly influences the nature of the targets of natural selection, the vectors of migration and the long-term retention of polymorphism, than for the genetic composition of populations.

Logically, genetic parameters reach their equilibrium value with lower temporal variation and faster than genotypic indices, even at small population sizes ($N \leq 1000$ in our simulations). Although they are poorly informative regarding $c$ below extreme values, accounting for genetic indices may limit the risk of misinterpretation when estimating $c$ not at equilibrium.

As a consequence, genotypic and genetic parameters appear to be complementary in terms of the estimation of $c$, with the former being helpful at equilibrium and for all values of $c$ (<0.95) and the latter being more accurate for estimating the incidence of clonality in populations not at equilibrium or discriminating among extreme values of $c$, which often implies a longer time needed to reach equilibrium (Reichel *et al.* 2016).

Unfortunately, these relationships cannot provide reliable information for detecting PC or estimating $c$ due to the pervasive effect of sampling on these parameters, which has proven particularly worrying and raises questions regarding many conclusions reached thus far in the literature as to the importance of sexual reproduction in a diverse range of partially clonal species.

*Detecting clonality under realistic conditions*

Based on our results, clonal richness ($R$) and clonal evenness (Pareto $\beta$) are highly sensitive to sampling. Even using relatively large sample sizes (from 100 to 500 individuals) leads to deeply biased estimates of the true $R$ and $\beta$ and thus $c$ values. $R$ is always greatly overestimated, by some orders of magnitude more than previously demonstrated with empirical datasets for which the rates of clonality remained unknown (Arnaud-Haond *et al.* 2007; Gorospe *et al.* 2015), and except in nearly strictly sexual populations, $\beta$ was also greatly overestimated (for $c \geq 0.1$). Genotypic descriptors computed from realistic sample sizes may be informative only for rare cases of small population sizes ($N \leq 1000$ individuals in the case of our simulations). For most situations where population sizes are large, genotypic descriptors computed with realistic sample sizes result in extreme underestimation of the rates of clonality (see below) or even in overlooking the occurrence of PC (*i.e.*, considering the species as strictly sexual). These results raise questions regarding the conclusions derived in the literature from studies assessing the occurrence or even sometimes the extent of clonality based only on genotypic indices.

In contrast, the distribution moments of $F_{IS}$ and mean LD for common sample sizes (more than 20 individuals) produced values consistent with those obtained from genotyping the whole population, yet they previously could be interpreted only for extreme rates of clonality (c≥0.95). Consequently, when analysing samples from populations with more than 1000 individuals, most genetic descriptors should remain informative and sometimes, together with any *R* values lower than 1, should be interpreted as a likely signature of a high prevalence of clonality (c≥0.95)

This worrying limitation recalls, for example, the results recently reported by Dia *et al.* (2014) for a unicellular phytoplankton species involved in harmful algal blooms (HABs), *Alexandrium minutum*. This species, which causes paralytic shellfish poisoning (PSP), shows an alternation between clonal (during the bloom) and sexual phases. Dia et al. (2004) sampled populations throughout the bloom (clonal) events, during which they grew from being nearly undetectable to exhibiting a concentration of $10^4$ to $10^5$ cells per litre. Of the more than 1000 strains cultivated, 265 were fully genotyped, among which no replicated genotypes were found, driving the estimate of clonal diversity to *R*=1. Without extensive knowledge of the biology of this species, clonality would not have been diagnosed on the basis of this sampling, which raises questions regarding the occurrence of clonality. Unfortunately, no $F_{IS}$ values could be reported in this study because only the haploid phase could be sampled, and the LD detected suggested the occurrence of recombination. However, according to these results, genetic descriptors allow the detection or estimation of clonality when its prevalence is extreme: the results by Dia *et al.* (2014) thus mainly suggest that the clonal rate during the bloom event did not exceed 0.95 in the few previous generations, still leaving great uncertainty as to the prevalence of sexual or clonal reproduction in this species.

Most target species in the literature, including clonal plants and invasive and pathogenic species, exhibit extremely large population sizes, thus raising serious questions regarding our ability to detect clonality based on realistic sample sizes, let alone infer its importance. The importance of sample size is reflected in the guidelines provided by the pioneering work of Tibayrenc *et al.* (1991), who listed 8 criteria to detect clonality, among which fixed heterozygosity, deviation from HWE and LD were expected to be important in the ability to diagnose clonality. Nevertheless, these criteria would apply only to diploid species with extreme rates of clonality, excluding haploid lineages and diploid species with $c<0.95$.

One may consider the clonal mechanisms and the way clonal replicates spatially disperse to better estimate the effect of the joint incidence of the sampling density and scale of dispersal of clones (driving the scale of spatial autocorrelation of genotypes compared to the grain size of sampling) on the ability of a given strategy to detect clonal replicates and therefore on the conclusions derived from population genetics data as to the incidence of sexual versus clonal reproduction. Along a continuum of dispersal from microorganisms such as unicellular algae and flying aphids to clonal plants with strong rhizomatic connections and ramets more often clumped than dispersed, the spatial autocorrelation of clones increases, as does the ability of a given sampling strategy to reveal clonal replicates at equal sampling densities. As a consequence, at the first end of this continuum, where spatial dispersal is not limited (as is the case for *A. minutum*), genotypic parameters alone may not be informative on the existence or extent of clonality except for nearly strictly clonal organisms such as the human pathogen *Trypanosoma cruzi*. Such power would be gained as the spatial distance of clonal dispersal becomes lower than the sampling mesh size (for an example of the influence of sampling strategy in corals, see

Gorospe *et al.* 2015; see Riginos 2015 for a comment), and clonal replicates would become decreasingly randomly diluted at large population sizes and across vast spatial scales.

*Quantifying clonality or merely evaluating its extent: how wrong can we be?*

In many studies, *R* may reflect the orders of magnitude separating sample size and population size (sometimes together with the clonal size and/or clumping of clonal replicates) rather than the prevalence of sexual reproduction. As illustrated in this work, even moderate values of *R* under our usually very small sampling densities (several tens of sampling units in populations bearing one hundred thousand to millions of them) may thus suggest a high prevalence of clonal reproduction. Some examples exist in the literature in which only a good knowledge of species biology prevents misleading conclusions based on values of genotypic diversity. These examples indicate the need to be very careful in interpreting genotypic parameters alone in the numerous cases where no such extensive knowledge of the species studied exists. An enlightening case is the study by Orantes *et al.* (2012) on aphids reproducing through cyclical parthenogenesis. Eight populations of *Aphis glycines* were sampled at two time steps corresponding to the early season, when sexual reproduction arises at rather small population sizes, and the late season, after a demographic explosion of populations under full clonality. Against all expectations based on a presumed relationship between *R* and *c* and ignoring the effect of sampling density, Orantes *et al.* (2012) found lower genotypic diversity during the season of sexual reproduction (average *R* of 0.85, average Pareto β of 2.9) than during the later season of pure clonality (average *R* of 0.97, average Pareto β of 4.2), *i.e.*, $R_{sexual} < R_{clonal}$ and $\beta_{sexual} < \beta_{clonal}$. Without knowledge of the cycle and a good understanding of the effect of population *versus* sample size, a higher rate of sexual reproduction in the late season could have been inferred. However, using the guidelines we aimed to develop here, *R* would mostly signal the significance of clonality and call for careful screening of genetic parameters. In fact, departure from HWE in this study confirms the complementarity of genotypic and genetic parameters by supporting the prevalence of clonal reproduction across the cycle, with mean $F_{IS}$ values of -0.21 and -0.24 in the earlier and later season, respectively, suggesting a more important influence of clonality in the later season, with a lower mean and larger variance. Similar patterns have been found in multiple studies on cyclical parthenogenetic species (e.g., Gilabert *et al.*, 2015; Loxdale *et al.* 2011). Another example is a highly clonal root-sucking nematode, *Xiphinema index*, which shows mid-range *R* values (0.16 to 0.39); however, negative mean $F_{IS}$ values with large variance in agreement with *LD* values suggest a rates of clonality exceeding 0.95 in all these populations, which had better agree with a naturalistic knowledge (Villate *et al.* 2010).

In fact, revising the numerous data acquired on clonal plants, including seagrasses, in light of the present results reveals very frequent negative $F_{IS}$ values, suggesting a much higher contribution of clonality than previously thought (Evans *et al.* 2014; Sinclair *et al.* 2014; Stoeckel *et al.* 2006) on the basis of their average *R* values (see Arnaud-Haond *et al.* 2019 for a meta-analysis). Unfortunately, $F_{IS}$ is often neglected in ecological studies, possibly due to difficulties in disentangling the influence of technical shortcomings such as null alleles from non-random mating such as selfing in some studies. In the seagrass literature, for example, moderate levels of *R* have led some authors to propose that sexual reproduction has a high incidence and may thus contribute greatly to recombination and dispersal through seed production (McMahon *et al.* 2017). The joint re-analysis of *R* and $F_{IS}$ values and their correlation can illuminate likely extreme but overlooked clonal rates (also see Arnaud-Haond *et al.* 2019). Although we seldom found this type of interpretation combining genotypic and genetic parameters in the literature (but see the examples above), this approach has been used by some authors, such as Ali *et al.* (2014) (also see the references above), to infer the importance of long-term clonality.

Interestingly, it has been shown that CloNcaSe, a method based on repeated genotypes alone (Ali *et al.* 2016), can deliver incorrect inferences, likely due to this subsampling effect on R. For a red alga (*Gracilaria chilensis*) maintained through strict clonality for generations, R values of 0.2 to 0.23 lead CloNcaSe to infer a ĉ=0.82, while ClonEstiMate, a second method based on transition probabilities of genotype frequencies (Becheler *et al.* 2017), correctly infers a ĉ=1. Similarly, an aphid population sampled when mostly clonal lineages can be found (*Rhopalosiphum padi,* Halkett *et al.* 2006) has an R of 0.89, leading CloNcaSe to infer a ĉ=0.68, while ClonEstiMate better inferred a ĉ=0.9.

Revising estimates of clonality in natural populations is particularly important because present-day interpretations, often mostly focusing on *R,* are likely to grossly underestimate its extent. A vast body of literature exists on the relationship between genotypic diversity and the resistance or resilience of populations, as demonstrated in experimental studies (Hughes *et al.* 2008; Hughes & Stachowicz, 2004; Reusch & Lampert, 2004; but see Massa *et al.* 2013). Severe overestimation of genotypic diversities may thus have led to strongly misleading conclusions as to the resilience of the studied populations, enhanced by their supposedly high *R* value, as well as to their ability to rely on dispersal of seeds due to recurrent events of sexual reproduction (Kendrick *et al.* 2017, 2012; McMahon *et al.* 2017). A case-by-case re-evaluation is thus needed to determine what may hold true for some species, depending on their life history traits (particularly longevity and turnover), but be completely incorrect for others.

*Conclusion*

To conclude, our results showed a large impact of PC on the genotypic composition of natural populations across the whole spectrum of all possible rates of clonality, supporting its strong influence on the tuning of evolutionary forces acting on these populations at different spatial and temporal scales, even at low values of *c*, as conjectured by Lewis (1987). By affecting the main path of emergence of new variants (somatic mutations rather than recombination), the targets of natural selection and migration ("… *the entity that persists and evolves is the clonal lineage…*"; Ayala, 1998), and the influence of drift (through the potentially much longer-term retention of polymorphism; Reichel *et al.* 2016; Yonezawa *et al.* 2004; and the present results), PC has the potential to profoundly influence both the short-term dynamics and the evolutionary trajectories of natural populations, even at a modest rate of clonality. Unravelling the occurrence of clonality and understanding its extent are thus of paramount importance for reconstructing, understanding and forecasting the demography, ecology and evolution of the vast number of (possibly including some that often remain undiagnosed) partially clonal species across the tree of life.

Unfortunately, given the present state of knowledge and existing analytical tools, the possibilities of inferring the rates of clonality using one episode of population genotyping are remote. These results also clarify the paradox of the often reported (but also often overlooked) combination of high genotypic diversities, suggesting both significant rates of sexual reproduction and significant heterozygote excess, supporting nearly strict clonality (Dia *et al.* 2014; Orantes *et al.* 2012). Many partially clonal organisms studied to date may rely on a much higher prevalence of clonal reproduction than initially thought, but clonal richness in these organisms may be overestimated due to the limited sampling power at hand. This work thus calls for a reappraisal of previously published data and conclusions on a broad range of clonal organisms. Perspectives on how to infer the importance of clonality using one episode of genotyping may, however, exist and can be summarised with the following guidelines:

1) PC can be detected or quantified with the usual sampling power and existing methods, mostly when the rate of clonality exceeds 95%.

2) Departure from HWE towards heterozygote excess, particularly together with a large variance in $F_{IS}$ across loci, indicates the occurrence and likely prevalence of clonality.

3) The joint examination of genotypic and genetic descriptors is often necessary when PC detection is still needed (a recommendation reminiscent of the ones formulated a long time ago for human pathogens (Tibayrenc *et al.* 1991; see also Tibayrenc and Ayala 2012) but seldom followed in ecological studies).

4) Considering both families of parameters may help better estimate the extent of clonal reproduction but may require accepting a large uncertainty, particularly when the rate of clonal reproduction is not very high.

5) As such departures are expected due to clonality, $F_{IS}$ should not be used

a- for the estimation of $p_{sex}$ (as initially offered by Douhovnikoff and Dodd 2003 and relayed by Arnaud-Haond *et al.* 2007), as it may be in most cases due to clonality rather than non-random pairing of gametes.

b- (perhaps not as strictly) when filtering next-generation sequencing (NGS) data based on possible PC. Such filters, failing to fit in the case of partial PC, would lead to at best a very large number of informative loci being discarded and at worst complete ignorance of the occurrence of PC in the dataset.

c- to detect technical artefacts such as null alleles and correct data or select loci using models based on pure sexuality, including those implemented in software, such as Micro-Checker (Van Osterhout *et al.* 2004).

6) Finally, due to the observed but faint signature of $c$ slightly below 95% in the second and further moments of $F_{IS}$ and to a lesser extent $r_d$, which remains visually undetectable but can be detected by machine learning methods, improvement is expected to result from using machine learning based on informed databases corresponding to the broadest possible range of scenarios. Such development represents a promising avenue and will require large and versatile databases to accommodate the diversity of life history traits associated with clonality and subsampling to account for sampling effects.

## Acknowledgements

This study greatly profited from the exchange of ideas within the CLONIX consortium. We warmly thank Fabien Halkett for his valuable comments that improved the clarity of the manuscript and Myriam Valero, the CLONIX consortium and particularly Ronan Becheler and Diane Bailleul for useful discussions.

This work was funded by the French National Research Agency (projects CLONIX: ANR-11-BSV7-007 and Clonix2D ANR-18-CE32-0001) and the INVAMAT project (Plant Health and Environment Division of the French National Institute of Agricultural Research).

This preprint has been reviewed and recommended by Peer Community In Evolutionary Biology (https://doi.org/10.24072/pci.evolbiol.100078). We thank Myriam Heuertz, Marcela Van Loo and David Macaya-Sanz for their constructive comments that enhanced the current version of our manuscript.

## Conflict of interest disclosure

The authors of this preprint declare that they have no financial conflicts of interest based on the content of this article. Sophie Arnaud-Haond is one of the PCI Ecology recommenders.

## Author contributions

SAH and SS laid the foundation of this work, conceived the study and wrote the manuscript. SS built the simulator and formalised the mathematical analyses and equations. SAH supervised BP's master training. BP, SAH and SS contributed to code enhancement, computation, data exploration and interpretation. BP tracked and managed the bibliography. All authors contributed to editing. SAH and SS were responsible for funding applications. All authors read and approved the final manuscript.

## References

Agapow, P. M., & Burt, A. (2001). Indices of multilocus linkage disequilibrium. *Molecular Ecology Notes, 1*(1-2), 101–102. doi: 10.1046/j.1471-8278.2000.00014.x.

Ali, S., Gladieux, P., Rahman, H., Saqib, M. S., Fiaz, M., Ahmad, H., de Vallavieille-Pope, C. (2014). Inferring the contribution of sexual reproduction, migration and off-season survival to the temporal maintenance of microbial populations: A case study on the wheat fungal pathogen *Puccinia striiformis f.sp. tritici. Molecular Ecology, 23*(3), 603–617. doi: 10.1111/mec.12629.

Ali, S., Soubeyrand, S., Gladieux, P., Giraud, T., Leconte, M., Gautier, A., . . . Enjalbert, J. (2016). Cloncase: Estimation of sex frequency and effective population size by clonemate resampling in partially clonal organisms. *Molecular Ecology Resources, 16*(4), 845–861. doi: 10.1111/1755-0998.12511.

Arnaud-Haond, S., Alberto, F., Teixeira, S., Procaccini, G., Serrão, E. A., & Duarte, C. M. (2005). Assessing genetic diversity in clonal organisms: Low diversity or low resolution? Combining power and cost efficiency in selecting markers. *Journal of Heredity, 96*(4), 434–440. doi: 10.1093/jhered/esi043.

Arnaud-Haond, S., Duarte, C., Alberto, F., & Serrao, E. (2007). Standardizing method to address clonality in population studies. *Molecular Ecology, 16*5, 115–5139. doi: 10.1111/j.1365-294X.2007.03535.x.

Arnaud-Haond, S., Stoeckel, S., Bailleul, D. (2019). New insights into the population genetics of partially clonal organisms: when seagrass data meet theoretical expectations. BioRxiv, . doi:

Avise, J., & Nicholson, T. (2008). *Clonality: The genetics, ecology, and evolution of sexual abstinence in vertebrate animals*. USA: Oxford University Press.

Avise, J. C. (2015). Evolutionary perspectives on clonal reproduction in vertebrate animals. *Proceedings of the National Academy of Sciences of the United States of America, 112*(29), 8867–8873. doi: 10.1073/pnas.1501820112.

Ayala, F. J. (1998). Is sex better? Parasites say "no". *Proceedings of the National Academy of Sciences of the United States of America, 95*(7), 3346–3348. doi: 10.1073/pnas.95.7.3346.

Bailleul D., Stoeckel, S., Arnaud-Haond, S. (2016). RClone: a package to identify MultiLocus Clonal Lineages and handle clonal data sets in R.- *Methods in Ecology and Evolution*, 7 (8), 966-970. doi: 10.1111/2041-210X.12550

Balloux, F., Lehmann, L., & de Meeûs, T. (2003). The population genetics of clonal and partially clonal diploids. *Genetics, 164*(4), 1635–1644.

Barrett, S. C. (2016). *Invasion genetics: The baker and stebbins legacy.* Chicester, West Sussex : John Wiley & Sons.

Barrett, S. C. H. (2015). Foundations of invasion genetics: The baker and stebbins legacy. *Molecular Ecology, 24*(9), 1927–1941. doi: 10.1111/mec.13014.

Becheler, R., Masson, J. P., Arnaud-Haond, S., Halkett, F., Mariette, S., Guillemin, M.-L., & Stoeckel, S. (2017). ClonEstiMate, a bayesian method for quantifying rates of clonality of populations genotyped at two-time steps. *Molecular Ecology Resources, 17*(6), e251–e267. doi: 10.1111/1755-0998.12698.

Berg, L.M., & Lascoux, M. (2000), Neutral genetic differentiation in an island model with cyclical parthenogenesis. *Journal of Evolutionary Biology*, *13*, 488-494. doi:10.1046/j.1420-9101.2000.00185.x

Bibr, C. (2018). Julie Sondra Decker, The Invisible Orientation: An Introduction to Asexuality. *Sexualities*, *21*(5–6), 840–842. https://doi.org/10.1177/1363460717724156

Decker, J.S. (2015). *The Invisible Orientation: An Introduction to Asexuality*. New York: Skyhorse Publishing. ISBN 9781634502436

De Meeûs, T., Lehmann, L., & Balloux, F. (2006). Molecular epidemiology of clonal diploids: A quick overview and a short DIY (do it yourself) notice. *Infection, Genetics and Evolution, 6*(2), 163–170. doi: 10.1016/j.meegid.2005.02.004.

De Meeûs, T., Prugnolle, F., & Agnew P. (2007) Asexual reproduction: Genetics and evolutionary aspects. *Cellular and Molecular Life Sciences, 64*, 1355–1372. doi: 10.1007/s00018-007-6515-2

Dia, A., Guillou, L., Mauger, S., Bigeard, E., Marie, D., Valero, M., & Destombe, C. (2014). Spatiotemporal changes in the genetic diversity of harmful algal blooms caused by the toxic dinoflagellate Alexandrium minutum. *Molecular Ecology, 23*(3), 549–560. doi: 10.1111/mec.12617.

Dorken, M. E., & Eckert, C. G. (2001). Severely reduced sexual reproduction in northern populations of a clonal plant, *Decodon verticillatus* (Lythraceae). *Journal of Ecology, 89*(3), 339–350. doi: 10.1046/j.1365-2745.2001.00558.x.

Douhovnikoff, V., & Dodd, R. S. (2003). Intra-clonal variation and a similarity threshold for identification of clones: Application to *Salix exigua* using AFLP molecular markers. *Theoretical and Applied Genetics, 106*(7), 1307–1315. doi: 10.1007/s00122-003-1200-9.

Eckert, C. G. (2002). The loss of sex in clonal plants. In J. F. Stuefer, B. Erschbamer, H. Huber, & J.-I. Suzuki (Eds.), *Ecology and evolutionary biology of clonal plants: Proceedings of clone-2000. An international workshop held in Obergurgl, Austria, 20–25 August 2000* (pp. 279–298). Dordrecht: Springer Netherlands.

Evans, S. M., Sinclair, E. A., Poore, A. G., Steinberg, P. D., Kendrick, G. A., & Vergés, A. (2014). Genetic diversity in threatened *Posidonia australis* seagrass meadows. *Conservation Genetics, 15*(3), 717–728. doi: 10.1007/s10592-014-0573-4.

Fehrer, J. (2010). Unraveling the mysteries of reproduction. *Heredity, 104,* 421–422. doi: 10.1038/hdy.2010.12.

Gilabert, A., Dedryver, C. A., Stoeckel, S., Plantegenest, M., & Simon, J. C. (2015). Longitudinal clines in the frequency distribution of 'super-clones' in an aphid crop pest. *Bulletin of Entomological Research, 105*(6), 694–703. doi: 10.1017/S0007485315000619.

Gorospe, K., Donahue, M., & Karl, S. A. (2015). The importance of sampling design: Spatial patterns and clonality in estimating the genetic diversity of coral reefs. *Marine Biology, 162*(5), 917–928. doi: 10.1007/s00227-015-2634-8.

Halkett, F., Kindlmann, P., Plantegenest, M., Sunnucks, P., & Simon, J. C. (2006). Temporal differentiation and spatial coexistence of sexual and facultative asexual lineages of an aphid species at mating sites. *Journal of Evolutionary Biology, 19,* 809–815.

Halkett, F., Simon, J. C., & Balloux, F. (2005). Tackling the population genetics of clonal and partially clonal organisms. *Trends in Ecology & Evolution, 20*(4), 194–201.

Hand, D. J., & Yu, K. (2001). Idiot's bayes: Not so stupid after all? *International Statistical Review / Revue Internationale de Statistique, 69*(3), 385–398. doi: 10.2307/1403452.

Hughes, A. R., Inouye, B. D., Johnson, M. T. J., Underwood, N., & Vellend, M. (2008). Ecological consequences of genetic diversity. *Ecology Letters, 11*(6), 609–623. doi: 10.1111/j.1461-0248.2008.01179.x.

Hughes, A. R., & Stachowicz, J. J. (2004). Genetic diversity enhances the resistance of a seagrass ecosystem to disturbance. *Proceedings of the National Academy of Sciences of the United States of America, 101*(24), 8998–9002. doi: 10.1073/pnas.0402642101.

Kendrick, G. A., Orth, R. J., Statton, J., Hovey, R., Montoya, L. R., Lowe, R. J., . . . Sinclair, E. A. (2017). Demographic and genetic connectivity: The role and consequences of reproduction, dispersal and recruitment in seagrasses. *Biological Reviews, 92*(2), 921–938. doi: 10.1111/brv.12261.

Kendrick, G. A., Waycott, M., Carruthers, T. J. B., Cambridge, M. L., Hovey, R., Krauss, S. L., & Verduin, J. J. (2012). The central role of dispersal in the maintenance and persistence of seagrass populations. *BioScience, 62*(1), 56–65. doi: 10.1525/bio.2012.62.1.10.

Lewis, W. M. (1987). The cost of sex. In S. C. Stearns (Ed.), *The evolution of sex and its consequences* (pp. 33–57). Basel: Birkhäuser Basel.

Liddell, H. G., Scott, R., Jones, H. S., & McKenzie, R. (1940). A Greek-English lexicon. Oxford: Clarendon Press.

Loxdale, H. D., Schöfl, G., Wiesner, K. R., Nyabuga, F. N., Heckel, D. G., & Weisser, W. W. (2011). Stay at home aphids: Comparative spatial and seasonal metapopulation structure and dynamics of two specialist tansy aphid species studied using microsatellite markers. *Biological Journal of the Linnean Society, 104*(4), 838–865. doi: 10.1111/j.1095-8312.2011.01761.x.

Marbà, N., & Duarte, C. M. (1998). Rhizome elongation and seagrass clonal growth. *Marine Ecology Progress Series,* 174, 269–280.

Marshall, D.R., & Weir B.S. (1979). Maintenance of genetic variation in apomictic plant populations. *Heredity*, *42*, 159–72.

Massa, S. I., Paulino, C. M., Serrão, E. A., Duarte, C. M., & Arnaud-Haond, S. (2013). Entangled effects of allelic and clonal (genotypic) richness in the resistance and resilience of experimental populations of the seagrass *Zostera noltii* to diatom invasion. *BMC Ecology, 13*(1), 39. doi: 10.1186/1472-6785-13-39.

McMahon, K. M., Evans, R. D., van Dijk, K. J., Hernawan, U., Kendrick, G. A., Lavery, P. S., & Waycott, M. (2017). Disturbance is an important driver of clonal richness in tropical seagrasses. *Frontiers in Plant Science, 8.* doi: 10.3389/fpls.2017.02026.

Navascués, M., Stoeckel, S., & Mariette, S. (2010). Genetic diversity and fitness in small populations of partially asexual, self-incompatible plants. *Heredity, 104,* 482–492. doi: 10.1038/hdy.2009.159.

Nougue, O., Rode, N.O., Jabbour-Zahab, R., Segard, A., Chevin, L.M., Haag, C.R. & Lenormand, T. (20015). Automixis in Artemia: solving a century-old controversy. *Journal of Evolutionary Biology, 28*(12), 2337-2348. doi: 10.1111/jeb.12757

Orantes, L. C., Zhang, W., Mian, M. A., & Michel, A. P. (2012). Maintaining genetic diversity and population panmixia through dispersal and not gene flow in a holocyclic heteroecious aphid species. *Heredity, 109*(2), 127–134. doi: 10.1038/hdy.2012.21.

Piñeiro, G., Perelman, S., Guerschman, J. P., & Paruelo, J. M. (2008). How to evaluate models: Observed vs. predicted or predicted vs. observed? *Ecological Modelling, 216*(3), 316–322. doi: 10.1016/j.ecolmodel.2008.05.006.

Putman, A. I., & Carbone, I. (2014). Challenges in analysis and interpretation of microsatellite data for population genetic studies. *Ecology and Evolution, 4*(22), 4399–4428. doi: 10.1002/ece3.1305.

Reichel, K., Masson, J. P., Malrieu, F., Arnaud-Haond, S., & Stoeckel, S. (2016). Rare sex or out of reach equilibrium? The dynamics of F IS in partially clonal organisms. *BMC Genetics, 17*(1), 76–76. doi: 10.1186/s12863-016-0388-z.

Reusch, T. B., & Lampert, W. (2004). *Role and maintenance of genetic diversity in natural populations.* Christian-Albrechts-Universität Kiel.

Riginos, C. (2015). Clones in space—how sampling can bias genetic diversity estimates in corals: Editorial comment on the feature article by Gorospe et al. *Marine Biology, 162,* 913–915. doi: 10.1007/s00227-015-2638-4.

Rouger, R., Reichel, K., Malrieu, F., Masson, J. P., & Stoeckel, S. (2016). Effects of complex life cycles on genetic diversity: Cyclical parthenogenesis. *Heredity, 117,* 336–347. doi: 10.1038/hdy.2016.52

Schön, I., Van Dijk, P., & Martens, K. (2009). *Lost sex: The evolutionary biology of parthenogenesis*. Dordrecht: Springer.

Sinclair, E., Krauss, S., Anthony, J., Hovey, R., & Kendrick, G. (2014). The interaction of environment and genetic diversity within meadows of the seagrass *Posidonia australis* (Posidoniaceae). *Marine Ecology Progress Series, 506,* 87–98. doi: 10.3354/meps10812.

Stoeckel, S., & Masson, J. P. (2014). The exact distributions of FIS under partial asexuality in small finite populations with mutation. *PLoS One, 9*(1), e85228. doi: 10.1371/journal.pone.0085228.

Stoeckel, S., Grange, J., Manjarres, J.F., Bilger, I., Frascaria-Lacoste, N., Mariette, S. (2006). Heterozygote excess in a self-incompatible and partially clonal forest tree species — Prunus avium L. *Molecular Ecology, 15*(8), 2109–18. doi: 10.1111/j.1365-294X.2006.02926.x

Tibayrenc, M., Avise, J. C., & Ayala, F. J. (2015). In the light of evolution IX: Clonal reproduction: Alternatives to sex. *Proceedings of the National Academy of Sciences of the United States of America, 112*(29), 8824–8826. doi: 10.1073/pnas.1508087112.

Tibayrenc, M., & Ayala, F. J. (2012). Reproductive clonality of pathogens: A perspective on pathogenic viruses, bacteria, fungi, and parasitic protozoa. *Proceedings of the National Academy of Sciences of the United States of America, 109*(48), E3305–E3313. doi: 10.1073/pnas.1212452109.

Tibayrenc, M., Kjellberg, F., Arnaud, J., Oury, B., Brenière, S. F., Dardé, M. L., & Ayala, F. J. (1991). Are eukaryotic microorganisms clonal or sexual? A population genetics vantage. *Proceedings of the National Academy of Sciences of the United States of America, 88*(12), 5129–5133. doi: 10.1073/pnas.88.12.5129.

Tibayrenc, M., Kjellberg, F., & Ayala, F. J. (1990). A clonal theory of parasitic protozoa: The population structures of entamoeba, giardia, leishmania, naegleria, plasmodium, trichomonas, and trypanosoma and their medical and taxonomical consequences. *Proceedings of the National Academy of Sciences of the United States of America, 87*(7), 2414–2418.

Van Oosterhout, C., Hutchinson, W. F., Wills, D. P., & Shipley, P. (2004). MICRO-CHECKER: Software for identifying and correcting genotyping errors in microsatellite data. *Molecular Ecology Notes, 4,* 535–538. doi: 10.1111/j.1471-8286.2004.00684.x

Villate, L., Esmenjaud, D., Van Helden, M., Stoeckel, S., & Plantard, O. (2010). Genetic signature of amphimixis allows for the detection and fine scale localization of sexual reproduction events in a mainly parthenogenetic nematode. *Molecular Ecology, 19*(5), 856–873. doi: 10.1111/j.1365-294X.2009.04511.x.

Webb, G. I., Boughton, J. R., & Wang, Z. (2005). Not so naive bayes: Aggregating one-dependence estimators. *Machine Learning, 58*(1), 5–24. doi: 10.1007/s10994-005-4258-6.

Weir, B. S., & Cockerham, C. C. (1984). Estimating F-statistics for the analysis of population structure. *Evolution, 38*(6), 1358–1370. doi: 10.2307/2408641.

Wright, S. (1921). Systems of mating. II. the effects of inbreeding on the genetic composition of a population. *Genetics, 6*(2), 124–143.

Wright, S. (1931). Evolution in mendelian populations. *Genetics, 16*(2), 97–159.

Wright, S. (1969). *Evolution and the genetics of populations, volume 2: Theory of gene frequencies*. Chicago: University of Chicago Press.

Yonezawa, K., Ishii, T., & Nagamine, T. (2004). The effective size of mixed sexually and asexually reproducing populations. *Genetics, 166*(3), 1529–1539.

Yu, F. H., Roiloa, S. R., & Alpert, P. (2016). Editorial: Global change, clonal growth, and biological invasions by plants. *Frontiers in Plant Science, 7*(1467). doi: 10.3389/fpls.2016.01467.

Zhang, H. (2004). *The optimality of naive bayes.* American Association for Artificial Intelligence.

**Supplementary material**

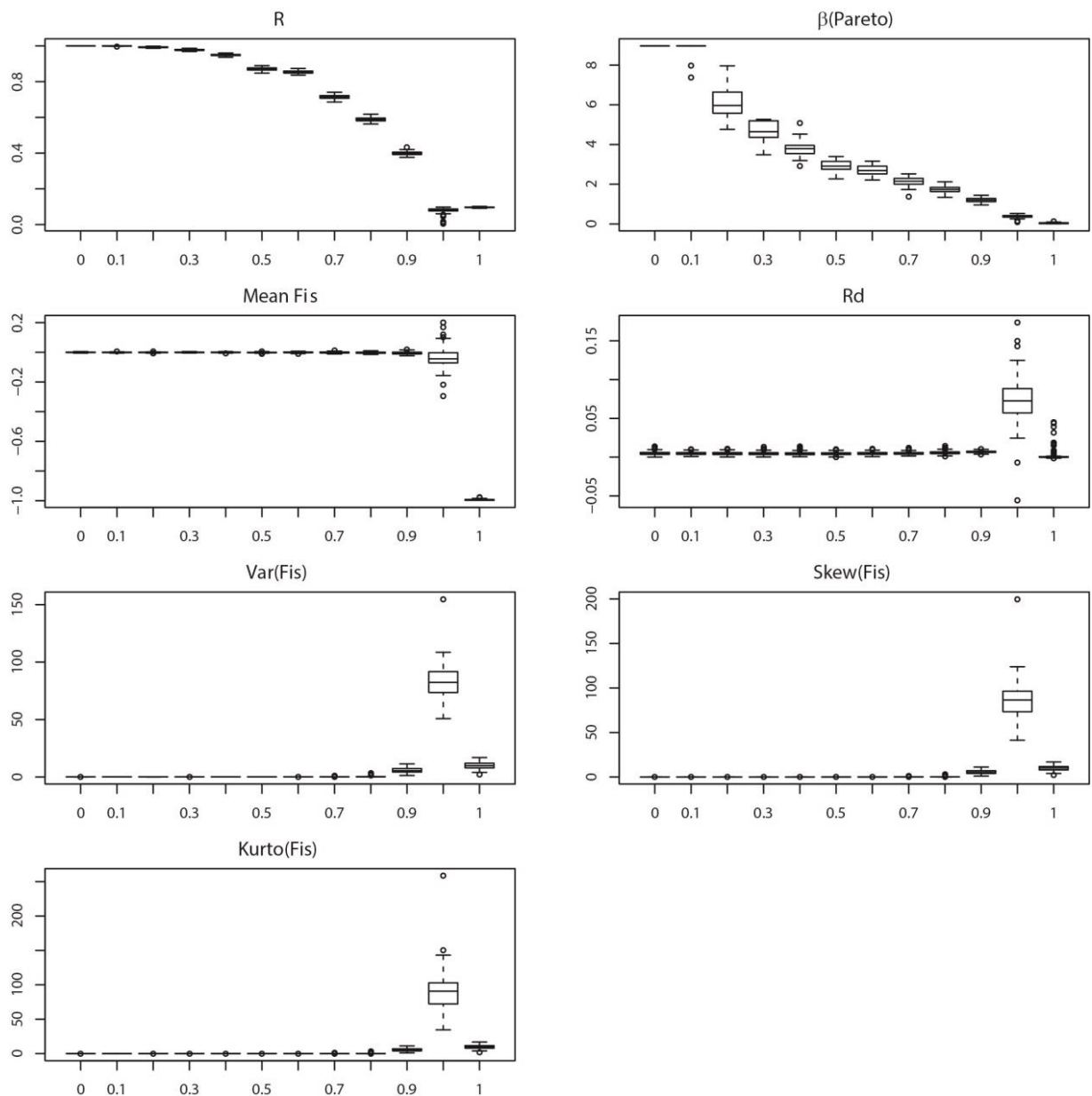

**Figure S1.a.**
Distribution of each parameter explored at equilibrium ($10^4$ generations of quantitatively homogeneous evolution since the initial random population): the genotypic parameters (a) R and (b) Pareto β and the genetic parameters (c) $F_{IS}$ mean, (d) $F_{IS}$ variance, (e) $F_{IS}$ skewness, (f) $F_{IS}$ kurtosis and (g) LD measured as $\bar{r}_d$ for increasing values of c at $N=10^3$. The X-axis is linear from c=0 to c=0.9 and then non-linear for the last two boxes at c=0.99 and c=1.

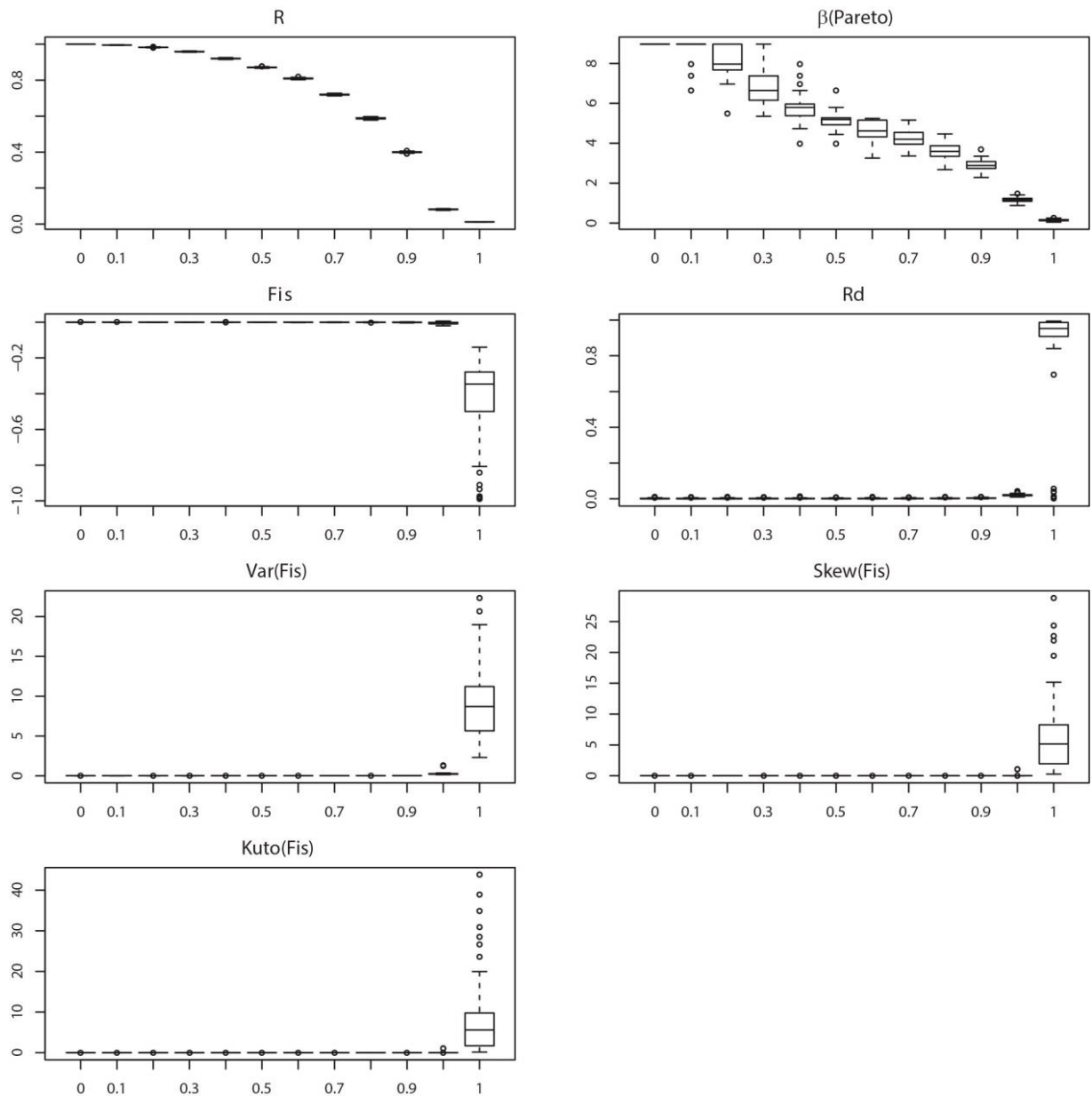

**Figure S1.b.**
Distribution of each parameter explored at equilibrium ($10^4$ generations of quantitatively homogeneous evolution since the initial random population): the genotypic parameters (a) R and (b) Pareto β and the genetic parameters (c) $F_{IS}$ mean, (d) $F_{IS}$ variance, (e) $F_{IS}$ skewness, (f) $F_{IS}$ kurtosis and (g) LD measured as $\bar{r}_d$ for increasing values of c at $N=10^4$. The X-axis is linear from c=0 to c=0.9 and then non-linear for the last two boxes at c=0.99 and c=1.

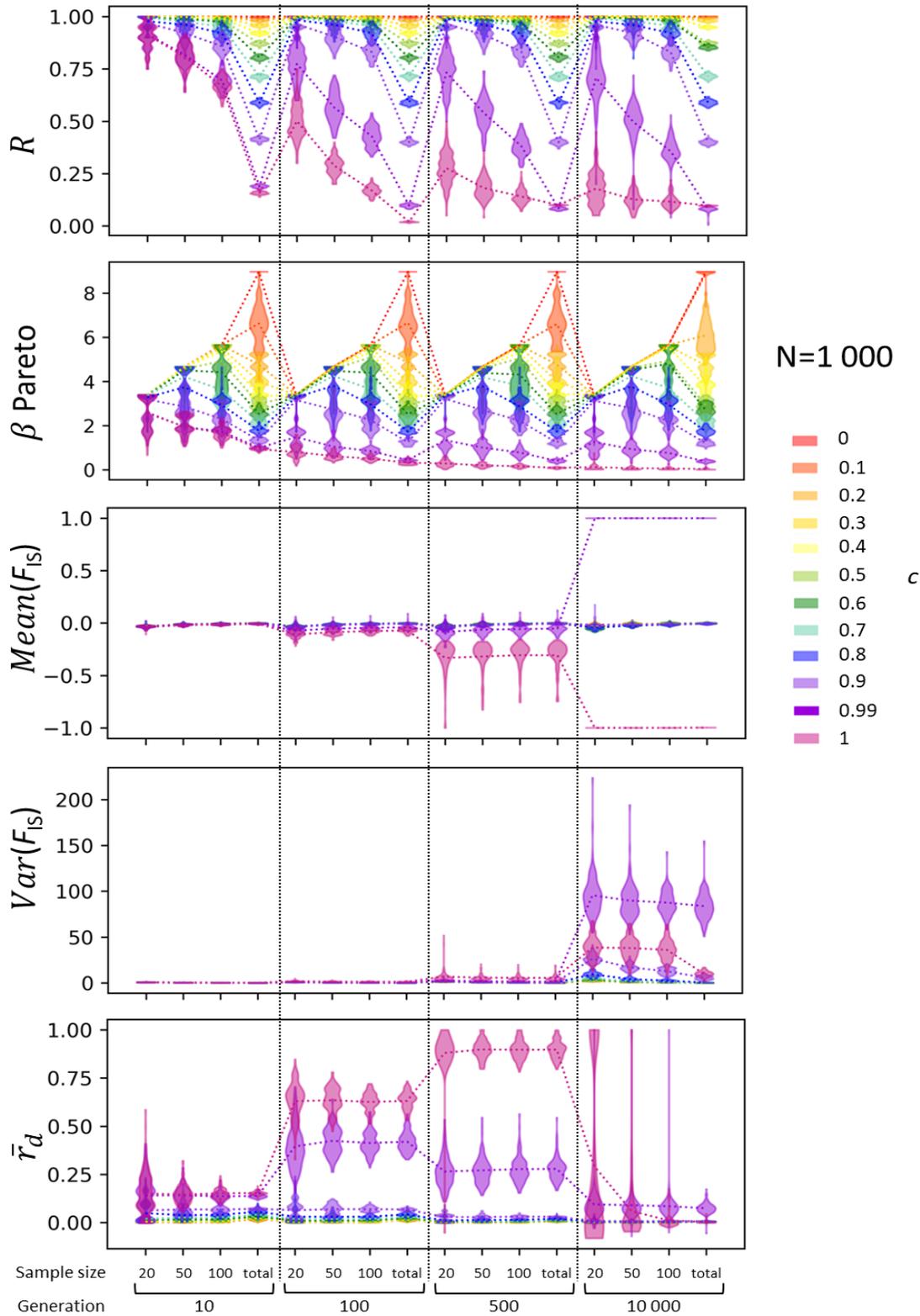

**Figure S2.a.** Temporal evolution of each parameter at a population size of $10^3$ individuals per generation as a function of the number of generations elapsed from a fully random population at generation 0 and sample sizes ($n_s$=20, 50, 100 and 1 000). Caution for interpretation: all x-axes are non-linear, and the y-axis for, $\bar{r}_d$ and the mean and variance of the $F_{IS}$ distributions present one to two changes in scaling.

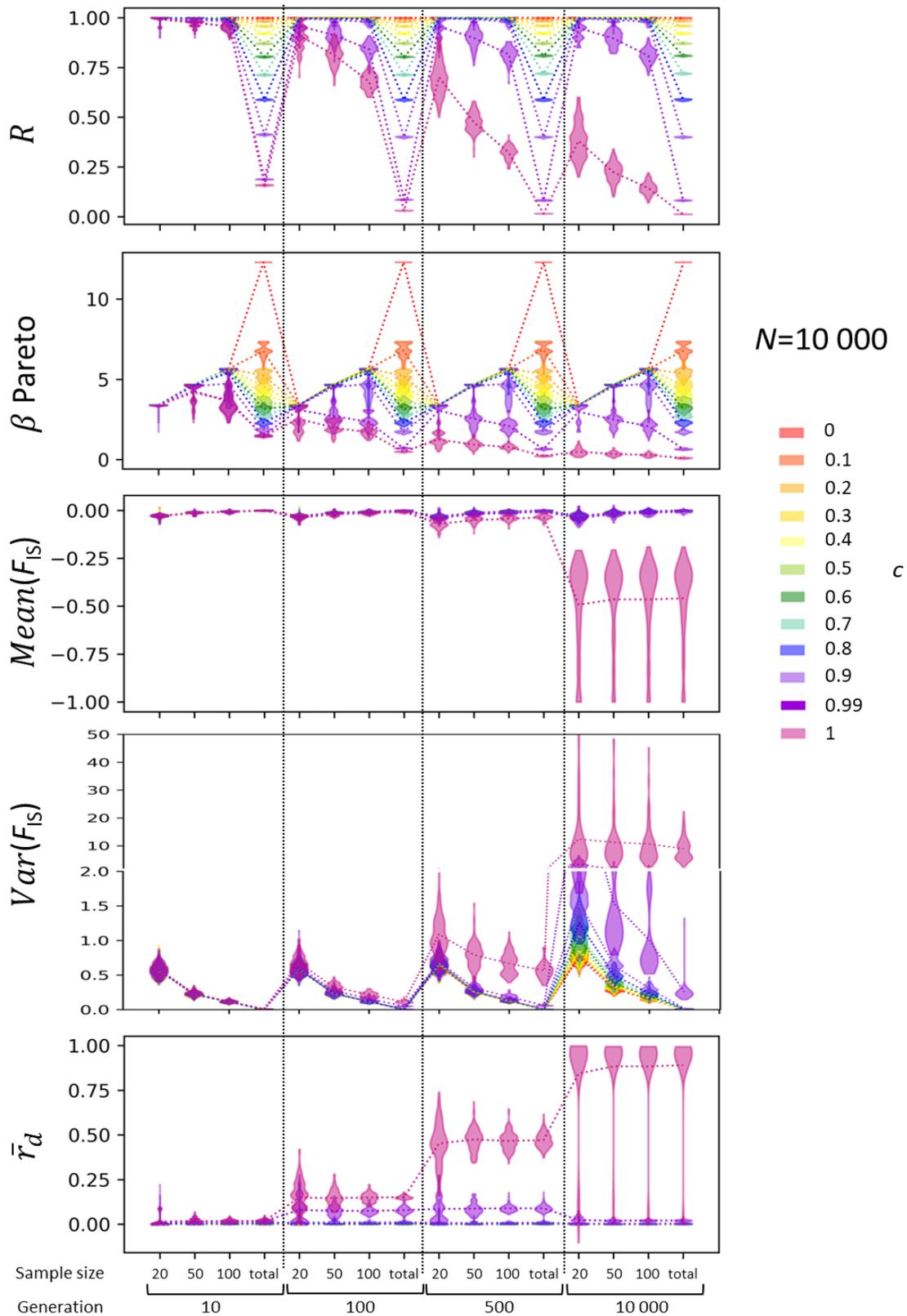

**Figure S2.b.** Temporal evolution of each parameter at a population size of $10^4$ individuals per generation as a function of the number of generations elapsed from a fully random population at generation 0 and sample sizes ($n_s$=20, 50, 100 and 10 000). Caution for interpretation: all x-axes are non-linear, and the y-axis for, $\bar{r}_d$ and the mean and variance of the $F_{IS}$ distributions present one to two changes in scaling.

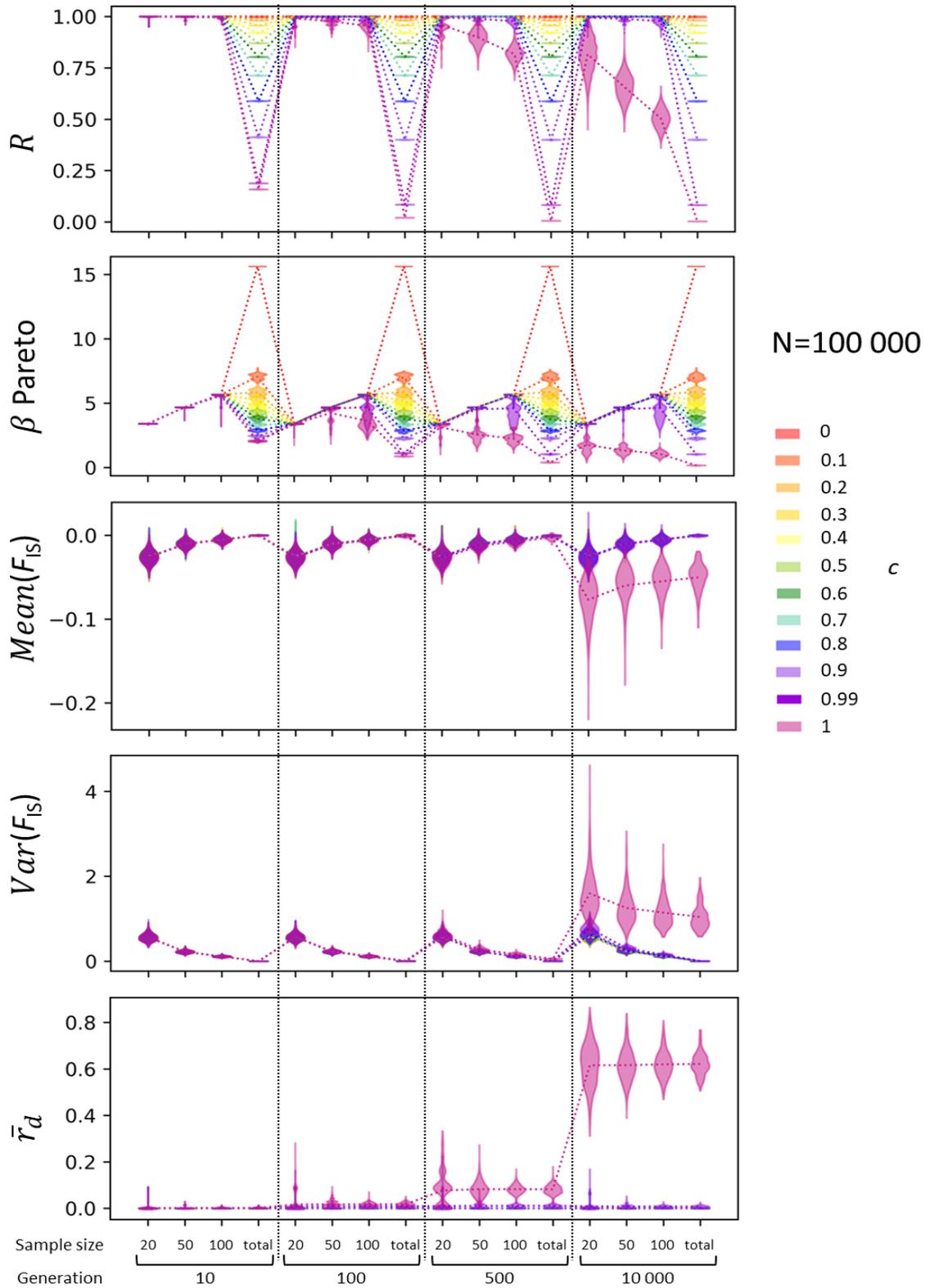

**Figure S2.c.** Temporal evolution of each parameter at a population size of $10^5$ individuals per generation as a function of the number of generations elapsed from a fully random population at generation 0 and sample sizes ($n_s$=20, 50, 100 and 100 000). Caution for interpretation: all x-axes are non-linear, and the y-axis for, $\bar{r}_d$ and the mean and variance of the $F_{IS}$ distributions present one to two changes in scaling.

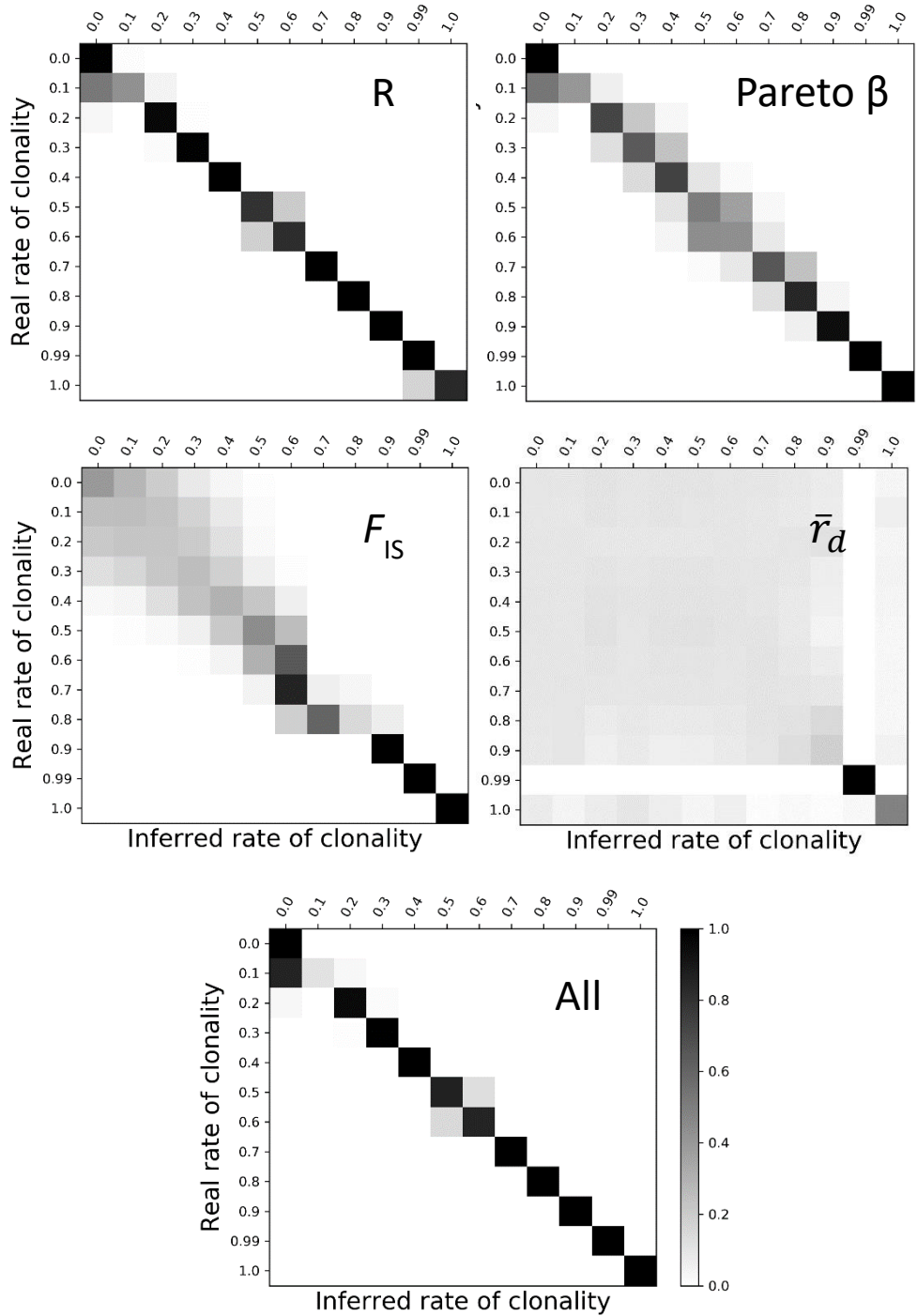

**Figure S3.a.** Supervised Bayesian inferences of $c$ at $N=10^3$ and for each parameter used for inference: the genotypic parameters (a) R and (b) Pareto β and the genetic parameters (c) $F_{IS}$ and (d) $\bar{r}_d$ as well as (e) the combination of all four parameters. The inferred values are plotted against the simulated values, with the density gradient from black to light grey indicating the most to least likely/probable.

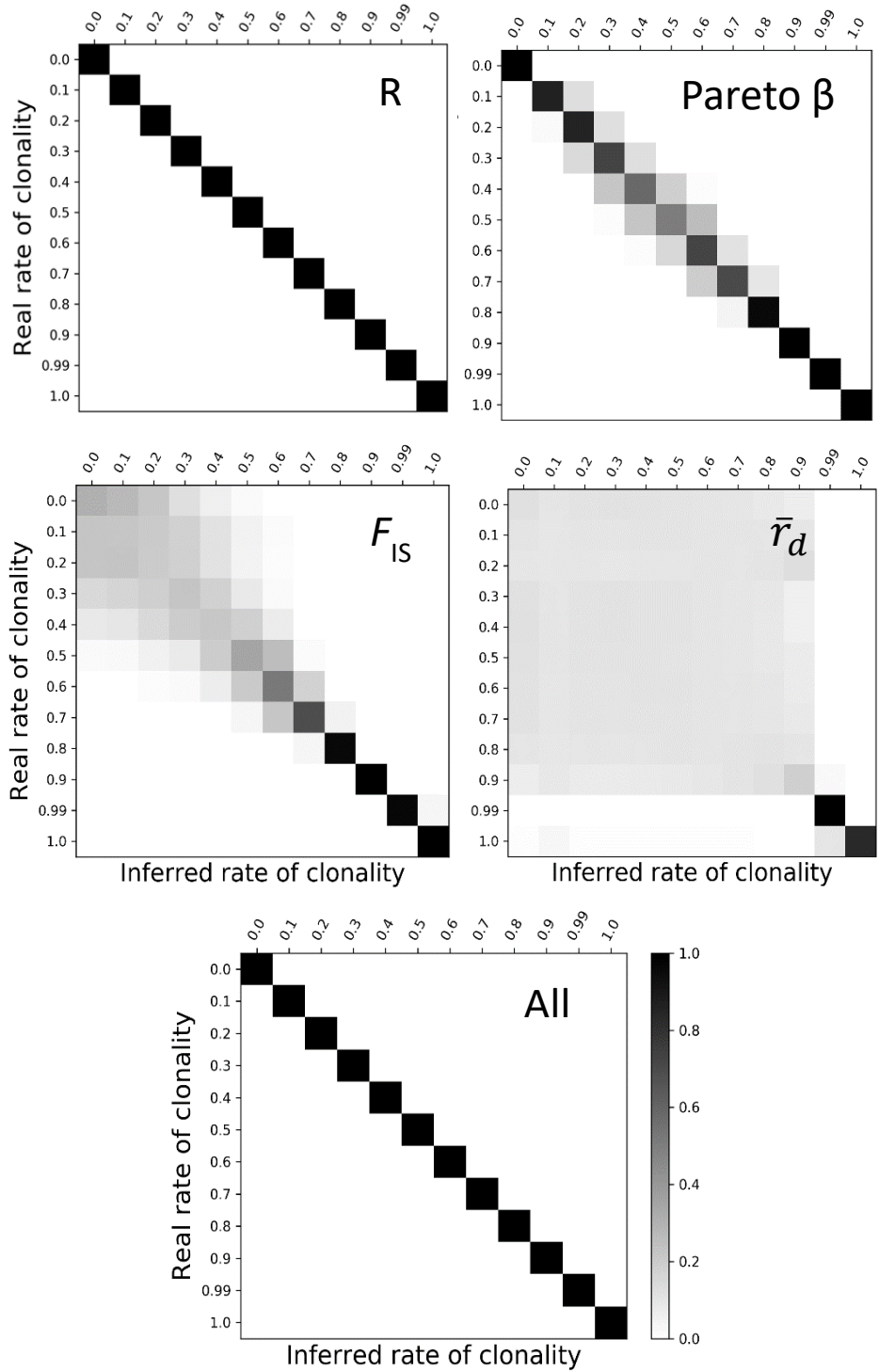

**Figure S3.b.** Supervised Bayesian inferences of *c* for N=10$^4$ and for each parameter used for the inference: the genotypic parameters (a) R and (b) Pareto β and the genetic parameters (c) $F_{IS}$ distribution, (d) $\bar{r}_d$ as well as (e) the combination of all four parameters. The inferred values are plotted against the simulated values, with the density gradient from black to light grey indicating the most to least likely/probable.

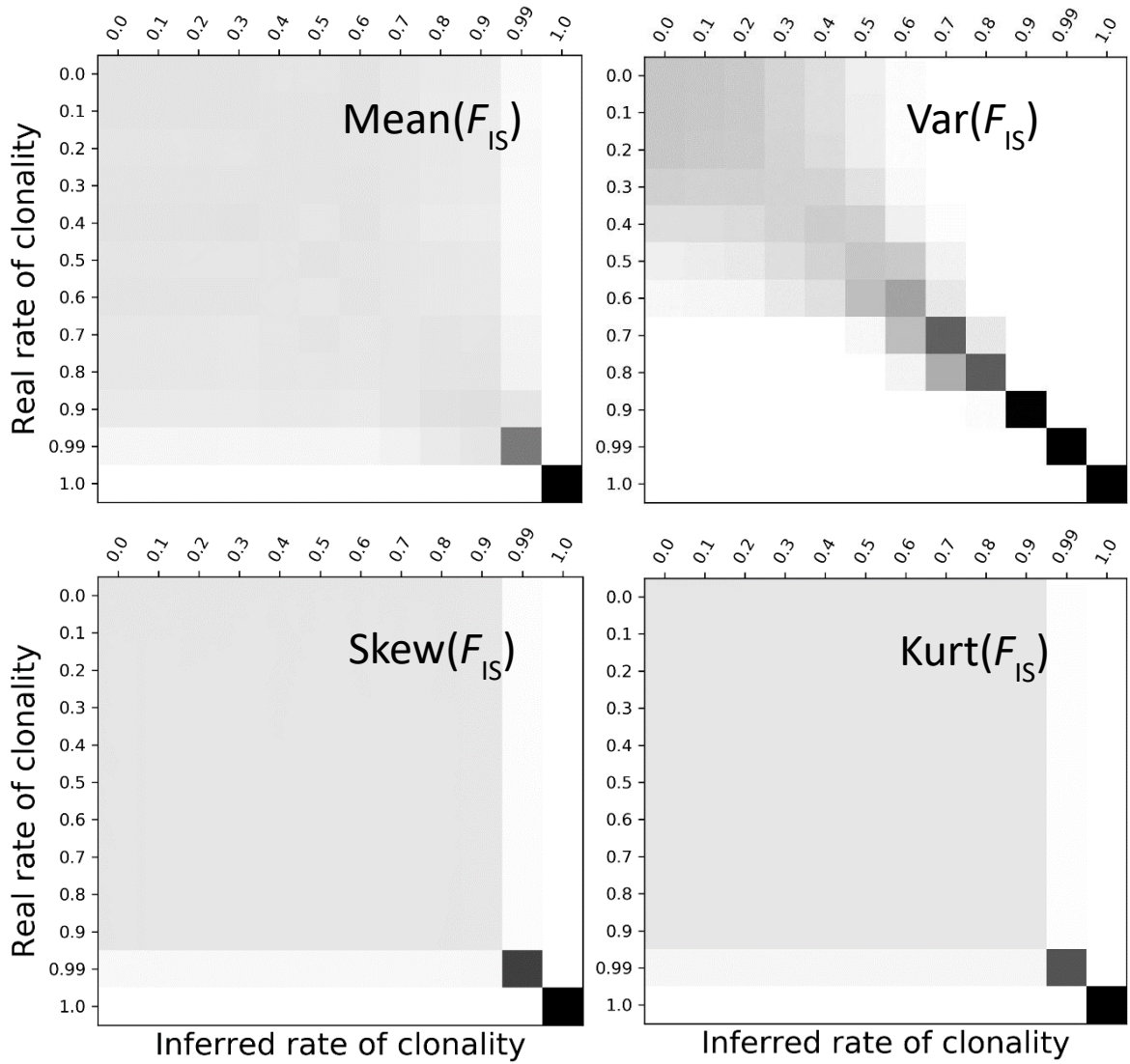

**Figure S4.** Supervised Bayesian inferences of *c* for N=10$^5$ using the four first moments of F$_{IS}$ distribution: (a) mean (Mean(F$_{IS}$)), (b) variance (Var(F$_{IS}$)), (c) skewness (Skew(F$_{IS}$)) and (d) kurtosis (Kurt(F$_{IS}$)). The inferred values are plotted against the simulated values, with the density gradient from black to light grey indicating the most to least probable.